\documentclass[12pt]{article} 
\usepackage[xdvi]{graphicx} 
\usepackage{latexsym}
\usepackage{amssymb}
\usepackage{amsmath} 
\usepackage{subfigure}
\usepackage{epsfig}
\usepackage{array}
\usepackage{cite} 
\usepackage{psfrag}
\usepackage{epsfig,t1enc,rotating}
\usepackage{a4,t1enc}

\begin{document}  

\newcommand{\todo}[1]{{\em \small {#1}}\marginpar{$\Longleftarrow$}}   
\newcommand{\labell}[1]{\label{#1}\qquad_{#1}} 

\thispagestyle{empty}
\vspace*{2.0cm}

\begin{center}
{\large \bf
Using D-Strings to Describe Monopole Scattering \\ - Numerical Calculations \\ 
}
\end{center}

\begin{center}
Jessica K. Barrett$^{a}$\footnote{jessica@raunvis.hi.is}, Peter Bowcock$^{b\,}$\footnote{peter.bowcock@durham.ac.uk}
\vspace{0.1cm}

\end{center}
\vspace*{1.0cm}

\centerline{$^a$ \it Science Institute} \centerline{\it University of Iceland} \centerline{\it Taeknigardi, Dunhaga 5}
\centerline{\it IS-107 Reykjavik, Iceland}

\bigskip\bigskip

\centerline{$^b$ \it Centre for Particle Theory} \centerline{\it
  Department of Mathematical Sciences} \centerline{\it University of
  Durham} \centerline{\it Durham, DH1 3LE, U.K.}

\bigskip\bigskip

\begin{abstract}
\nocite{Barrett:2004zt} 
We calculate the energy radiated during the scattering of two D-strings stretched between two D3-branes, working from the Born-Infeld action for the D-strings. The ends of the D-strings are magnetic monopoles from the point of view of the gauge theory living on the D3-branes, and so the scattering we describe is equivalent to monopole scattering. Our results suggest that no energy is radiated during the scattering, in contrast to the monopole result of ref. \cite{Manton:1988bn}.

\end{abstract}

\newpage \baselineskip=18pt \setcounter{footnote}{0} 
\setcounter{page}{1}

\section{Introduction}

It has been known for some time that a D-string ending on a D3-brane looks like a magnetic monopole from the point of view of the gauge theory living on the D3-brane. In ref. \cite{Callan:1998kz} it was shown that a fundamental string ending on a D-brane can be described as a solitonic solution of the Born-Infeld action of the D-brane. This led to some speculation as to whether a D-string ending on a D-brane can be described in a similar way. In refs. \cite{Constable:1999ac} and \cite{Constable:2001ag} it was shown that this is indeed the case for a D-string ending on a D3-brane (ref. \cite{Constable:2001ag} also contains a discussion of the D5-brane case). This D-brane configuration has been widely studied in the literature. See also, for example refs. \cite{Diaconescu:1997rk}, \cite{Hashimoto:1997px}, \cite{Ghoroku:1999bc}, \cite{Brecher:1998tv} and \cite{Gauntlett:1999xz}, and refs. \cite{Hamanaka:2001pe} and \cite{Rey:1998ik} for related work.

In ref. \cite{Barrett:2004zt} we initiated a calculation of the energy radiated during the scattering of two D-strings stretched between D3-branes, using the D-strings' action. Our aim was to compare our result with the monopole result, which was shown by Manton and Samols in ref. \cite{Manton:1988bn} to be
\begin{equation}
E_{rad} \sim 1.35 \, m_{mon} v_{-\infty}^5 \,\,\, , \quad \frac{E_{rad}}{E_{tot}} \sim 1.35\, v_{-\infty}^3 \,\,\, ,
\label{eq:Erad}
\end{equation}
where $E_{rad}$ is the energy radiated, $E_{tot}$ is the total energy in the system, $m_{mon}$ is the mass of each monopole and $v_{-\infty}$ is the asymptotic velocity of each monopole. In ref. \cite{Barrett:2004zt} we discussed the soliton solution to the D-string Born-Infeld action that represents two D-strings stretched between two D3-branes. We used this solution to describe D-string scattering using the moduli space approximation of ref. \cite{Manton:1982mp}. Then we calculated equations of motion for perturbations to the moduli space approximation, since the perturbations contain the information about the energy radiated. Previous work regarding perturbations of the BIon spike can be found in refs. \cite{Lee:1997xh}, \cite{Kastor:1999ag}, \cite{Savvidy:1999wx} and \cite{Rey:1997sp}.  

In this paper we will conclude our investigation with our numerical calculations for the energy radiated during D-string scattering. We will seek to solve numerically the full equations of motion, rather than using the moduli space approximation. 

The layout of the paper is as follows. In section \ref{sec:review} we will review the D3-D1 brane configuration, including the results of our first paper, ref. \cite{Barrett:2004zt}. We will discuss some of the implications for our numerical calculations, and make some points about the energy of the system. In section \ref{sec:asymptotic_calculations} we will describe the motion of the D-strings in the asymptotic limit, when they are far apart. We will discuss how the solutions can be split into zero modes and non-zero modes, and we will show that the non-zero modes decouple from the zero modes in this limit. It is the energy in the non-zero modes after scattering that represents the energy radiated. In section \ref{sec:numerics} we will describe our numerical calculations to solve the equations of motion, and in section \ref{sec:energy-radiated} we will use these numerical solutions to calculate the energy radiated. Section \ref{sec:conclusions} contains our conclusions.

\section{Describing Magnetic Monopoles Using the Born-Infeld Action}
\label{sec:review}

In this section we will briefly review some of the background material, and discuss the implications for our calculations

\subsection{Monopoles as Soliton Solutions of the Born-Infeld Action}

We start by reviewing the description of magnetic monopoles as soliton solutions of Born-Infeld actions from refs. \cite{Constable:1999ac} and \cite{Constable:2001ag}. The relevant D-brane configuration is D-strings attached to D3-branes. Note that this configuration can be studied either using the D3-brane Born-Infeld action, or using the D-strings' Born-Infeld action. 

Consider the Born-Infeld action for a D3-brane with the magnetic field on the brane, $B_i$, excited, and with a single transverse field, $\Phi$, excited. If we look for a static solution which minimises the energy, we find that the solution obeys the usual BPS equations for a magnetic monopole
\begin{equation}
B_i = D_i \Phi \,\, ,
\label{eq:BPS}
\end{equation}
where $D_i$ is the covariant derivative with respect to the gauge field on the D3-brane and $i,j=1,2,3$ label the spacelike dimensions of the D3-brane. The simplest solution to \eqref{eq:BPS} is
\begin{equation}
\Phi(r) = \frac{N}{2r} \,\, , \quad \vec{B}(\vec{r}) = \mp \frac{N}{2r^3} \vec{r} \,\, ,
\end{equation}
where $r$ is the radial coordinate in the D-brane's worldvolume. Note that this solution for $\Phi(r)$ indicates that the D3-brane has been pulled into an infinitely long spike in the direction corresponding to the field $\Phi$. Calculating the energy and Ramond-Ramond charge of this solution, we can conclude that it represents $N$ semi-infinite D-strings attached to the D3-brane at the origin.

On the other hand, this configuration can also be studied using the non-Abelian Born-Infeld action for $N$ D-strings. See refs. \cite{Tseytlin:1997cs}, \cite{Tseytlin:1999dj} and \cite{Myers:1999ps} for more details of the non-Abelian Born-Infeld action, and ref. \cite{Brecher:2005sj} for a recent discussion. Exciting three transverse scalars in the action, say $\Phi^1$, $\Phi^2$ and $\Phi^3$, and looking for a soliton solution, leads to the BPS equations
\begin{equation}
\label{eq:nahm}
\partial_{\sigma} \Phi^i  + \frac{1}{2i} \, \epsilon_{ijk} \, [\Phi^j,\Phi^k] = 0 \,\, ,
\end{equation} 
 where $\sigma$ is the D-strings' spatial direction. The equations \eqref{eq:nahm} are Nahm's equations from the ADHMN construction of a magnetic monopole (see ref. \cite{Corrigan:1984sv} for a review of the ADHMN construction, and refs. \cite{Tsimpis:1998zh} and \cite{Chen:2002vb} for discussions of how the ADHMN boundary conditions apply in this case). The solution corresponding to $N$ semi-infinite D-strings `funnelling out' into a D3-brane is
 \begin{equation}
 \Phi^{i} = \pm \frac{\alpha^i}{2\sigma} \,\, ,
 \end{equation}
 where the $\alpha^i$ are an $N \times N$ representation of the $SU(2)$ algebra. 
 
 We next review the solution to Nahm's equations which describes two D-strings stretched between two D3-branes, which we discussed in our previous paper ref. \cite{Barrett:2004zt}. We defined a new string coordinate, $\xi = 2\sigma/L$, where $L$ is the distance between the D3-branes (note that this coordinate transformation sets the distance between the D3-branes to be 2). We introduced the ansatz
 \begin{equation}
 \Phi^i = \frac{2}{L}  f_i(\xi,t) \, \sigma_i  \quad \textrm{no summation over $i$} \,\, ,
 \end{equation}
 where $\sigma_i$ are the Pauli matrices. Then Nahm's equations reduce to
 \begin{equation}
 f_1' - f_2 f_3 =0 \,\, , \quad f_2' - f_3 f_1 =0 \,\, , \quad f_3' - f_1 f_2 =0 \,\, ,
 \label{eq:bog}
 \end{equation}
 where $'$ denotes differentiation with respect to $\xi$. The appropriate solutions to these equations, which were first derived in ref. \cite{Brown:1982gz}, are
 \begin{eqnarray}
f_{1}(\xi,k) & = & \frac{-K(k)}{\textrm{sn}(K(k)\xi,k)} \,\, , \quad
f_{2}(\xi,k)  =  \frac{-K(k)\textrm{dn}(K(k)\xi,k)}{\textrm{sn}(K(k)\xi,k)} \,\, , \nonumber \\
\label{eq:fsoln}
f_{3}(\xi,k) & = & \frac{-K(k)\textrm{cn}(K(k)\xi,k)}{\textrm{sn}(K(k)\xi,k)} \,\, ,
 \end{eqnarray} 
 where $K(k)$ is the complete elliptic integral of the first kind and $\text{sn}(\xi,k)$, $\text{cn}(\xi,k)$ and $\text{dn}(\xi,k)$ are the Jacobian 
elliptic functions. (See ref. \cite{Erdelyi} for a review of the properties of elliptic functions.) The parameter $k$ is a modulus with $0 \leq k < 1$. The $f_i$ all have poles at $\xi=0$ and $\xi=2$, which correspond to the D-strings `funnelling out' into D3-branes. Also, symmetry properties of the Jacobian elliptic functions imply that $f_1$ and $f_2$ are symmetric about $\xi=1$, while $f_3$ is antisymmetric.

\subsection{Describing Monopole Scattering}
\label{sec:scattering}

In ref. \cite{Barrett:2004zt} we explained how to describe the scattering of two D-strings using the solutions \eqref{eq:fsoln}. 

In the limit $k \to 1$ the $f_i$ flatten out, and we can write
\begin{equation}
f_1 \sim - K(k) \,\, , \quad f_2 \sim 0 \,\, , \quad f_3 \sim 0 \,\, , \quad \textrm{as $k \to 1$} \,\, .
\label{eq:f_kto1}
\end{equation} 
These approximations are accurate, except near $\xi=0$ and $\xi=2$, where the $f_i$ have poles. This limit corresponds to the two D-strings being far apart on the $x_1$-axis, with positions $x_1 = \pm f_1$.

In order to describe the scattering we take the $f_i$ with $k$ close to 1 as an initial condition. We start the motion by allowing $k$ to depend on time, $k \to k(t)$, such that initially the D-strings are moving slowly towards each other. Then $k$ decreases with time until it reaches $k=0$, when the configuration is axially symmetric in the $x_1$-$x_2$ plane, since $f_1(\xi,0)=f_2(\xi,0)$. At this point $f_1$ and $f_2$ swap roles, and $k$ increases towards 1, so that the D-strings are moving apart along the $x_2$-axis, and have therefore scattered at $90^{\circ}$.

Note that the description of scattering in the previous paragraph relied on the moduli space approximation of ref. \cite{Manton:1982mp}; the solutions were only allowed to depend on time through $k(t)$. In the moduli space approximation the motion of the D-strings follows a geodesic in moduli space, so that at any point in time the solution has the form of the static soliton solutions \eqref{eq:fsoln}. The D-strings have the same velocity at the end of scattering as they did initially, and the potential energy is always constant, so that no energy is radiated in this approximation. To calculate the energy radiated we have to take into account higher order corrections to the moduli space approximation. We can write the full solutions to the equations of motion as
\begin{equation}
\varphi_i(\xi,t)=f_i(\xi,k(t)) + \epsilon_i(\xi,t) \,\, ,
\label{eq:varphi}
\end{equation}
where the motion of the D-strings has been split into the zero modes, the $f_i$, and the non-zero modes, the $\epsilon_i$. The zero modes behave according to the moduli space approximation, and the non-zero modes act as small perturbations to the zero modes (we can assume that the $\epsilon_i$ are small when the D-strings are moving slowly). After the D-strings have scattered, the energy in the non-zero modes represents the energy radiated during scattering, as we will discuss in section \ref{sec:decoupling} below.

\subsection{Yang-Mills vs. Born-Infeld}

In order to calculate the energy radiated numerically we will work with the Born-Infeld action in the low energy limit, $\alpha' \to 0$. There are two ways to take this limit (see ref. \cite{Barrett:2004zt}). The distance between the D3-branes is given by
\begin{equation}
L = \alpha' v \,\, ,
\end{equation}
where, in the D3-brane description, $v$ is the expectation value of the field $\Phi$, which plays the role of the Higgs field. So when we take the limit $\alpha' \to 0 $ we can either keep $v$ fixed, in which case $L\to 0$, or we can keep $L$ fixed, in which case $v \to \infty$.

The mass of the monopole/D-string is given by
\begin{equation}
m_{mon} = T_1 L= \frac{v}{g_s} \,\, ,
\end{equation}
where $T_1$ is the tension of the D-string, and $g_s$ is the string coupling. Therefore the appropriate limit to take is $\alpha' \to 0$, $v$ fixed, so that the mass of the monopole is finite. Unfortunately, taking this limit leads to an action with very complicated equations of motion, which we were unable to solve numerically. We will therefore take the alternative limit, in which $L$ is fixed, and $v \to \infty$. The resulting action is a Yang-Mills action
\begin{equation}
S_{YM} \sim \int_{-\infty}^{\infty} \!\! dt \int_0^2 d\xi \left( \dot{\varphi}_1^2 + \dot{\varphi}_2^2 + \dot{\varphi}_3^2 - (\varphi'_1 - \varphi_2 \varphi_3)^2 - (\varphi'_2 - \varphi_3 \varphi_1)^2   - (\varphi'_3 - \varphi_1 \varphi_2)^2   \right) \,\, ,
\label{eq:S_YM}
\end{equation}
where $\Phi_i = 2 \varphi_i \sigma_i/L$. The BPS equations for this action are identical in form to those derived from the full action, equation \eqref{eq:bog},
\begin{equation}
\varphi'_1 - \varphi_2 \varphi_3 = 0 \,\, , \quad \varphi'_2 - \varphi_3 \varphi_1 = 0 \,\, , \quad \varphi'_3 - \varphi_1 \varphi_2 = 0 \,\, .
\label{eq:bog_varphi}
\end{equation} 
The equations of motion are
\begin{eqnarray}
\label{eq:phi1eqn}
\ddot{\varphi}_1 - \varphi''_1 + \varphi_1(\varphi_2^2 + \varphi_3^2) & = & 0  \,\, ,\\
\label{eq:phi2eqn}
\ddot{\varphi}_2 - \varphi''_2 + \varphi_2(\varphi_3^2 + \varphi_1^2) & = & 0  \,\, ,\\
\label{eq:phi3eqn}
\ddot{\varphi}_3 - \varphi''_3 + \varphi_3(\varphi_1^2 + \varphi_2^2) & = & 0  \,\, .
\end{eqnarray}
It is these equations that we have solved numerically, as we will describe in section \ref{sec:numerics}. 

Since the mass of the D-strings is infinite in the limit we are taking, the energy radiated during scattering will also be infinite, since it is proportional to $m_{mon}$. To keep our calculations finite we will calculate the ratio of the energy radiated to the total energy (which is conserved - see section \ref{sec:energy-cons}), since this quantity will not depend on $m_{mon}$.

\subsection{Energy Considerations}

In this section we outline some general points about the energy of the Yang-Mills system we wish to solve.

\subsubsection{The Energy Densities}
\label{sec:energy-densities}

The kinetic and potential energy densities of the system governed by the Yang-Mills action \eqref{eq:S_YM} are
\begin{eqnarray}
\text{K.E. density} & = & \frac{T}{2} \left( \dot{\varphi}_1^2 + \dot{\varphi}_2^2 + \dot{\varphi}_3^2 \right) \,\, ,
\label{eq:KE-density} \\
\text{P.E. density} &  = & \frac{T}{2} \left( (\varphi'_1 - \varphi_2\varphi_3)^2 + (\varphi'_2 -
\varphi_1\varphi_3)^2 + (\varphi'_3 - \varphi_1\varphi_2)^2 \right)\,\, ,
\label{eq:PE-density}
\end{eqnarray}
where $T$ is a constant which is determined by the mass of the monopole. We will not need an exact expression for $T$ since we will always be dealing with ratios of energies.

Note that the potential energy density for a solution obeying the BPS equations \eqref{eq:bog_varphi} is zero, as we would expect. The potential energy density of a solution therefore measures the deviation of the solution away from the BPS solution.

\subsubsection{Energy Conservation}
\label{sec:energy-cons}

We will show here that the total energy remains conserved. We consider the 
Noether currents for time translation $t \to t-a$, which are 
\begin{eqnarray}
j^0 & = & \frac{T}{2} \left( (\dot{\varphi}_1^2 + \dot{\varphi}_2^2 + \dot{\varphi}_3^2) +
(\varphi_1' - \varphi_2 \varphi_3)^2 + (\varphi_2' - \varphi_3 \varphi_1)^2 + (\varphi_3' - \varphi_1 \varphi_2)^2 \right)  
\nonumber \,\, , \\
j^1 & = & T \big( \varphi'_1\dot{\varphi}_1 + \varphi'_2\dot{\varphi}_2 + \varphi'_3\dot{\varphi}_3 - \partial_t (\varphi_1 \varphi_2 \varphi_3) \big)
\nonumber \,\, .
\end{eqnarray}
The total energy $E_{tot}$ is given by
\begin{displaymath}
E_{tot} = \int_0^2 \! j^0 d\xi \,\, .
\end{displaymath}
By current conservation we have
\begin{equation}
\dot{E}_{tot}  =  \int_0^2 \! \partial_{\xi} \, j^1d\xi \,\, ,
\end{equation}
which gives
\begin{equation}
\dot{E}_{tot}  = T \, [\varphi'_1\dot{\varphi}_1 + \varphi'_2\dot{\varphi}_2 + \varphi'_3\dot{\varphi}_3- \partial_t (\varphi_1 \varphi_2 \varphi_3)]^2_0 \,\, .
\label{eq:dtE}
\end{equation}
Using equations \eqref{eq:varphi-series} of appendix \ref{sec:decoupling_proof} we have 
\begin{eqnarray}
\varphi'_1\dot{\varphi}_1 + \varphi'_2\dot{\varphi}_2 + \varphi'_3\dot{\varphi}_3 - \partial_t (\varphi_1 \varphi_2 \varphi_3) & = &  O(\xi) \nonumber \\
 & = & 0 \quad \text{at $\xi=0$} \,\, .
 \end{eqnarray}
As discussed below equation \eqref{eq:varphieqn-1} in appendix \ref{sec:decoupling_proof}, $\varphi_1$ and $\varphi_2$ will be evolved in such a way that they are symmetric about $\xi=1$, and $\varphi_3$ is antisymmetric. The contribution to \eqref{eq:dtE} at $\xi=2$ is therefore also
zero. So we have
\begin{displaymath}
\dot{E}_{tot} = 0 \,\, ,
\end{displaymath}
the total energy of the system is conserved.

\section{The D-Strings' Motion in the Asymptotic Limit}
\label{sec:asymptotic_calculations}

In this section we discuss the motion of the two D-strings in the limit when they are very far apart and moving very slowly. In section \ref{sec:zero_mode} we will calculate a series expansion for the position of the D-strings in this limit, which agrees with the monopole calculation  of ref. \cite{Manton:1988bn}. In section \ref{sec:decoupling} we will discuss the decoupling of the D-strings' motion zero modes and non-zero modes in this limit.  

\subsection{The Zero-Mode Motion}
\label{sec:zero_mode}

When the D-strings are very far apart their interaction is minimal, and we can neglect the contributions from the non-zero modes. The D-strings being very far apart corresponds to the limit $k \to 1$ in the solutions \eqref{eq:fsoln} (see section \ref{sec:scattering}). In this limit we can expand these solutions as series in $k' \equiv \sqrt{1 - k^2}$. We find
\begin{align}
f_1(\xi,k)  =  -\frac{K(k)}{\sinh(\xi K(k))} & \bigg( \cosh(\xi K(k)) + {\displaystyle\frac{1}{4} \frac{{\displaystyle \xi K(k)}}{{\displaystyle\sinh(\xi K(k))}}(k')^2} - {\displaystyle\frac{1}{4} \cosh(\xi K(k))(k')^2}  \nonumber \\ \label{eq:f1-kseries}
& \quad + \cdots \bigg) \,\, ,
\end{align} 
and
\begin{align}
f_2(\xi,k)  =  -\frac{K(k)}{\sinh(\xi K(k))} & \bigg({\displaystyle1 + \frac{1}{4} \frac{\xi K(k) \cosh(\xi K(k))}{\sinh(\xi K(k))}(k')^2 + {\displaystyle \frac{1}{4} \cosh^2(\xi K(k))(k')^2} }\nonumber \\ \label{eq:f2-kseries}
& \quad {\displaystyle - \frac{(k')^2}{2}} + \cdots\bigg) \,\, ,  
\end{align} 
and
\begin{align}
f_3(\xi,k)  =  -\frac{K(k)}{\sinh(\xi K(k))} & \bigg({\displaystyle1 + \frac{1}{4} \frac{\xi K(k) \cosh(\xi K(k))}{\sinh(\xi K(k))}(k')^2 - \frac{1}{4} \cosh^2(\xi K(k))(k')^2} \nonumber \\ \label{eq:f3-kseries}
& \quad + \cdots \bigg) \,\, .
\end{align} 
(see Appendix \ref{sec:k-series} for the details of this calculation). The parameter $k$ is unnatural to work with because the solutions for $f_1$, $f_2$ and $f_3$ depend on it in a highly nonlinear fashion. We can use instead $K(k)$, which gives the approximate position of the D-strings in the $x_1$-direction when they are very far apart (see equation \ref{eq:f_kto1}). This is possible in the asymptotic limit because the expansions \eqref{eq:f1-kseries} - \eqref{eq:f3-kseries} only depend on $k$ through $K(k)$. The expansions for $f_1$, $f_2$ and $f_3$ in terms of $K$ are best expressed as series in $e^{-2K}$. We find
\begin{align}
f_1(\xi,K) = - \frac{K}{\sinh(\xi K)} \,\, & \bigg( {\displaystyle \cosh(\xi K) + 4\left(\frac{\xi K}{\sinh(\xi K)} - \cosh(\xi K)\right) e^{-2K} } \nonumber \\
& \quad + O(K^2 e^{-4K}) \Big) \,\, ,
\label{eq:f1-Kseries}
\end{align}
and
\begin{align}
f_2(\xi,K) = - \frac{K}{\sinh(\xi K)} \,\, & \bigg( {\displaystyle 1 + 4\left( \frac{\xi K\cosh(\xi K)}{\sinh(\xi K)} + \cosh^2(\xi K) - 2\right) e^{-2K} } \nonumber \\ \label{eq:f2-Kseries}
& \quad + O(K^2 e^{-4K}) \bigg) \,\ ,
\end{align}
and
\begin{align}
f_3(\xi,K) = - \frac{K}{\sinh(\xi K)} \,\, & \bigg( {\displaystyle 1 + 4\left(\frac{\xi K\cosh(\xi K)}{\sinh(\xi K)} - \cosh^2(\xi K) \right) e^{-2K} } \nonumber \\ \label{eq:f3-Kseries}
& \quad + O(K^2 e^{-4K}) \bigg) \,\, ,
\end{align}
(see Appendix \ref{sec:K-series}).

In the moduli space approximation the D-strings' motion is described by allowing the modulus $K$ to depend on time. We assume that $\dot{K}$ is small so that the D-strings are moving towards each other slowly (this is necessary for the moduli space approximation to be accurate). Energy conservation now takes the form
\begin{equation}
\label{eq:energy-conservation}
\frac{T}{2} \int_0^2 \left(\frac{df_1}{dt}\right)^2 + \left(\frac{df_2}{dt}\right)^2 + \left(\frac{df_3}{dt}\right)^2 d\xi = E \,\, ,
\end{equation}
where we have neglected the contribution of the potential energy because it is very small in the asymptotic limit. Here $E$ is a constant which represents the initial energy of the D-strings, i.e. their energy when they are an infinite distance apart. Writing \eqref{eq:energy-conservation} in terms of the parameter $K(t)$, and using that $f_1$, $f_2$ and $f_3$ only depend on time through $K$, we obtain
\begin{equation}
\label{eq:energy-conservation-1}
\frac{T}{2} \, \dot{K}^2 \int_0^2 \left(\frac{df_1}{dK}\right)^2 + \left(\frac{df_2}{dK}\right)^2 + \left(\frac{df_3}{dK}\right)^2 d\xi = E \,\, .
\end{equation}
Then the expression for $\dot{K}$ is given by
\begin{equation}
\label{eq:Kdot}
\dot{K} = \sqrt{\frac{2E}{I_1 T}}\,\, ,
\end{equation}
where
\begin{equation}
I_1 = \int_0^2 \left(\frac{df_1}{dK}\right)^2 + \left(\frac{df_2}{dK}\right)^2 + \left(\frac{df_3}{dK}\right)^2 d\xi \,\, .
\label{eq:I1}
\end{equation}
Using the expansions \eqref{eq:f1-Kseries} - \eqref{eq:f3-Kseries} for $f_1$, $f_2$ and $f_3$ in \eqref{eq:I1} we obtain
\begin{equation}
I_1 = 2 \left( 1 - \frac{1}{K} +O(Ke^{-2K}) \right) \,\, .
\label{eq:I1_expanded}
\end{equation}
We differentiate \eqref{eq:energy-conservation-1} to obtain an expression for $\ddot{K}$
\begin{equation}
\ddot{K} = -E \, \frac{I_2}{I_1^2} \,\, ,
\end{equation}
where
\begin{equation}
I_2 = \int_0^2 \left(\frac{df_1}{dK}\right)\left(\frac{d^2f_1}{dK^2}\right) + \left(\frac{df_2}{dK}\right)\left(\frac{d^2f_2}{dK^2}\right) + \left(\frac{df_3}{dK}\right)\left(\frac{d^2f_3}{dK^2}\right) d\xi \,\, . 
\label{eq:I2}
\end{equation}
Again using the expansions \eqref{eq:f1-Kseries} - \eqref{eq:f3-Kseries} in \eqref{eq:I2}, we get
\begin{equation}
I_2 = \frac{1}{K^2} \left(1 + O(K^3 e^{-2K}) \right) \,\, .
\end{equation}

The leading order terms in the expression \eqref{eq:Kdot} give (using equation \eqref{eq:I1_expanded})
\begin{equation}
\label{eq:Kdot-leading}
\dot{K} = |v_{-\infty}| \left( 1 - \frac{1}{K} \right)^{-1/2} \,\, .
\end{equation}
where $v_{-\infty}$ is the velocity of the D-strings in the asymptotic limit $t \to -\infty$. Expanding equation \eqref{eq:Kdot-leading} as a series in $1/K$, and integrating gives
\begin{equation}
\label{eq:K-expansion}
K - \frac{1}{2} \ln K + \frac{1}{8K} + O\left(\frac{1}{K^2}\right) = v_{-\infty}(t+t_0) \,\, ,
\end{equation}
where $t_0$ is a constant parameter corresponding to the freedom to translate the problem in time. To first order the solution to \eqref{eq:K-expansion} is
\begin{equation}
\label{eq:K-firstorder}
K = v_{-\infty}(t+t_0) \,\, .
\end{equation}
We can find higher order solutions by perturbing \eqref{eq:K-firstorder}, substituting the perturbed solution back into \eqref{eq:K-expansion}, and solving for the perturbation. The resulting expression for $K(t)$ is
\begin{eqnarray}
K & = & v_{-\infty}(t+t_0) + \frac{1}{2} \ln(v_{-\infty}(t+t_0)) + \frac{\ln(v_{-\infty}(t+t_0))}{4v_{-\infty}(t+t_0)} - \frac{1}{8v_{-\infty}(t+t_0)} \nonumber \\
&& + O\left(\frac{(\ln(v_{-\infty}t)^2}{(v_{-\infty}t)^2} \right) \,\, .
\label{eq:K_expansion}
\end{eqnarray} 
\begin{sloppy}
Note that \eqref{eq:K_expansion} is in agreement with the equivalent expression from the three-dim\-en\-sion\-al monopole calculation, given in equation (8) of ref. \cite{Manton:1988bn}. 
\end{sloppy}

\subsection{Decoupling of Zero Modes and Non-Zero Modes}
\label{sec:decoupling}

In section \ref{sec:scattering} we described how the D-strings' motion can be thought of as being split into two parts; the motion of the zero modes, i.e. the motion of the centres of mass of the D-strings, and the motion of the non-zero modes, which act as perturbations on top of the zero modes. Energy can be transferred between the zero modes and the non-zero modes as a result of the interaction between the two D-strings. But when they are far apart, and the interaction can be neglected, the zero modes and the non-zero modes decouple. We give an argument to show the decoupling between zero modes and non-zero modes in appendix \ref{sec:decoupling_proof}. This means that energy can no longer be transferred between zero modes and non-zero modes, as we show explicitly in appendix \ref{sec:energy_decoupling}. It is the energy which has been transferred between zero modes and non-zero modes as a result of D-string scattering that represents the energy radiated during scattering.

\section{Solving the Equations of Motion Numerically}
\label{sec:numerics}

In this section we will describe the numerical methods we used to solve the Yang-Mills equations of motion, \eqref{eq:phi1eqn} - \eqref{eq:phi3eqn}. In section \ref{sec:numerical_methods} we will discuss the numerical methods we used. In section \ref{sec:initial_conditions} we will specify the initial conditions we used, and in section \ref{sec:boundary_conditions} we will discuss our boundary conditions. In section \ref{sec:phi_results} we will present some graphs of our results for the $\varphi_i$.

\subsection{Numerical Methods} 
\label{sec:numerical_methods}

Note from equations \eqref{eq:varphi-series} of Appendix \ref{sec:decoupling_proof} that near $\xi = 0$ the $\varphi_i$ take the form 
\begin{equation}
\varphi_i(\xi,t) = -\frac{1}{\xi} + O(\xi) \,\, .
\end{equation}
So the $\varphi_i$ have singularities at $\xi=0$, but the singularities are constant in time. In order to handle these singularities numerically we removed them by defining the fields $g_i$ as follows
\begin{equation}
\label{eq:gi}
g_i(\xi,t) \equiv \frac{1}{\xi} + \varphi_i(\xi,t) \,\, .
\end{equation}
This also implies that
\begin{equation}
\label{eq:gidot}
\dot{g}_i(\xi,t) = \dot{\varphi}_i(\xi,t) \,\, .
\end{equation}
The equations of motion for the fields $g_i$ are
\begin{align}
\label{eq:g1eqn}
\ddot{g}_1 - g''_1 + g_1(g_2^2+g_3^2) - \frac{(g_2^2+g_3^2+2g_1g_2+2g_1g_3)}{\xi}
+ \frac{(2g_1+2g_2+2g_3)}{\xi^2} & =  0  \,\, , \\  
\label{eq:g2eqn}
\ddot{g}_2 - g''_2 + g_2(g_3^2+g_1^2) - \frac{(g_3^2+g_1^2+2g_2g_3+2g_2g_1)}{\xi}
+ \frac{(2g_1+2g_2+2g_3)}{\xi^2} & =  0  \,\, , \\  
\label{eq:g3eqn}
\ddot{g}_3 - g''_3 + g_3(g_1^2+g_2^2) - \frac{(g_1^2+g_2^2+2g_3g_1+2g_3g_2)}{\xi}
+ \frac{(2g_1+2g_2+2g_3)}{\xi^2} & =  0  \,\, .  
\end{align}
The third and fourth terms in these equations are apparently singular at $\xi=0$. However, using again the series solutions for the $\varphi_i$, equation \eqref{eq:varphi-series} from appendix \ref{sec:decoupling_proof}, we have
\begin{eqnarray}
g_1(\xi,t) & = & a_1\xi + O(\xi^3)  \,\, , \quad
g_2(\xi,t)  =  b_1\xi + O(\xi^3) \,\, , \nonumber \\
\label{eq:g-series}
g_3(\xi,t) & = & c_1\xi + O(\xi^3) \,\, , 
\end{eqnarray}
with
\begin{equation}
a_1 + b_1 + c_1 = 0 \,\, .
\end{equation}
From this we can deduce that the terms which appear to be singular in \eqref{eq:g1eqn} - \eqref{eq:g3eqn} are in fact finite at $\xi=0$.

The $\varphi_i$ also have singularities at $\xi=2$, but since we will only solve numerically for the range $0<\xi<1$, these singularities will not affect our numerical calculations. 

We solved the equations numerically by evolving an initial configuration in time using an RK4 procedure, adapted to two-dimensional partial differential equations, subject to certain boundary conditions (see sections \ref{sec:initial_conditions} and \ref{sec:boundary_conditions} respectively for discussions of the initial conditions and boundary conditions). A detailed discussion of the method we used is given in ref. \cite{piette}. Our program was adapted from the RK4 program given in ref. \cite{piette}, and used routines from ref. \cite{Press:1992} to calculate the Jacobian elliptic functions numerically. Since the effect we were seeking to observe was so small, it was necessary to achieve very accurate results. We did this by using 7 points for the calculation of $g_i''$ in the RK4 method. We also used very small stepsizes; $d\xi=0.0001$ as the spatial stepsize, and $dt=0.00005$ as the time stepsize.

\subsection{Initial Conditions}
\label{sec:initial_conditions}

We start the motion at $t=0$ with the monopoles moving tangential to the static
solutions, so
\begin{equation}
\label{eq:varphi-initial-condition}
\varphi_i(\xi,0) = f_i(\xi,k_0) \,\, ,
\end{equation}
where $k_0$ is chosen so that the D-strings are sufficiently far apart initially. In \cite{Manton:1988bn}, Manton and Samols found that two monopoles cease to interact with one another for $r > 10$. So we take $k_0 = 0.9999999999$, for which $K(k_0) = 12.55264624 \approx f_1(\xi,0)$.

At $t=0$ the D-strings should be moving towards each other very slowly, so we set
\begin{equation}
\label{eq:varphidot-initial-condition}
\dot{\varphi}_i(\xi,0) = \dot{k}_0 \frac{df_i}{dk} \bigg|_{(\xi,k_0)} \,\, ,
\end{equation} 
where $\dot{k}_0$ is fixed by the initial velocity of the D-strings as follows. Since $f_1$ is approximately constant in $\xi$ initially, and $f_2$ and  
$f_3$ are approximately zero, $\dot{f}_1(1,k_0)$ is a good approximation to the initial velocity, $v_{init}$ of the D-strings (recall $0<\xi<2$, so $\xi=1$ is the midpoint of the strings). So, having specified $v_{init}$, we can calculate $\dot{k}_0$ from the following equation
\begin{equation}
\label{eq:initial-kdot}
v_{init} = \dot{f}_1(1,k_0) = \dot{k}_0 \frac{df_1}{dk}\bigg|_{(1,k_0)} \,\, .
\end{equation}

The initial conditions for the $g_i$ and $\dot{g}_i$ can be deduced from the initial conditions for the $\varphi_i$ and $\dot{\varphi}_i$ respectively using the definitions \eqref{eq:gi} and \eqref{eq:gidot}. 

Note that the initial configurations for $\varphi_1$, $\varphi_2$ ($\varphi_3$) and $\dot{\varphi}_1$, $\dot{\varphi}_2$ ($\dot{\varphi}_3$) are symmetric (antisymmetric) about $\xi=1$, because the $f_i$ have these symmetry properties. As we discuss in appendix \ref{sec:decoupling_proof}, the $\varphi_i$ will be evolved in time in such a way that these symmetries are perserved. We therefore used these symmetries to reduce the numerical computation by solving the equations of motion for the range $0 < \xi <1$, and transforming these solutions appropriately about $\xi=1$ to obtain solutions for the full range of $\xi$.

\subsection{Boundary Conditions}
\label{sec:boundary_conditions}

We consider first the left-hand border, $\xi=0$. The series expansions \eqref{eq:g-series} for the $g_i$ near $\xi=0$ imply
\begin{equation}
\label{eq:lh-boundary-condition}
g_i(0,t) = \dot{g}_i(0,t) = 0 \,\,.
\end{equation}
For points close to the left-hand border it was not possible to use 7 points to calculate $g_i''$ in the RK4 evolution; we used instead a 3-point calculation for the point next to the border, and a 5-point calculation for the next point along.

To fix the right-hand border we used the symmetry properties of the $\varphi_i$ about $\xi=1$. These symmetries imply that
\begin{equation}
\varphi'_1(1,t)  =  \varphi'_2(1,t) = 0 \,\, , \quad
\varphi_3(1,t)  =  0 \,\, .
\end{equation}
These imply for the $g_i$
\begin{eqnarray}
\label{eq:RHg1g2}
g'_1(1,t) & = & g'_2(1,t) = -1 \,\, , \\
\label{eq:RHg3}
g_3(1,t) & = & 1 \,\, .
\end{eqnarray}
On the right-hand border $g_3$ is therefore fixed, and we have
\begin{equation}
g_3(1,t) = 1 \,\, , \quad \dot{g}_3(1,t) = 0 \,\, .
\end{equation}
However, $g_1(1,t)$ and $g_2(1,t)$ vary in time, and so we used an RK4 procedure to calculate them. We calculated $g_1''$ and $g_2''$ for the RK4 method using the symmetry of $\varphi_1$ and $\varphi_2$ about $\xi=1$, and using 5 points. In the same way we calculated $g_1''$ and $g_2''$ for the two points next to the right-hand border.

\subsection{Results for the $\varphi_i$}
\label{sec:phi_results}

\begin{figure}
\centering
\psfrag{xi}{{\scriptsize$\xi$}}
\psfrag{phi1}{{\footnotesize$\varphi_1$}}
\psfrag{phi2}{{\footnotesize$\varphi_2$}}
\psfrag{phi3}{{\footnotesize$\varphi_3$}}
\mbox{
      \subfigure[$\varphi_1$]{\epsfig{figure=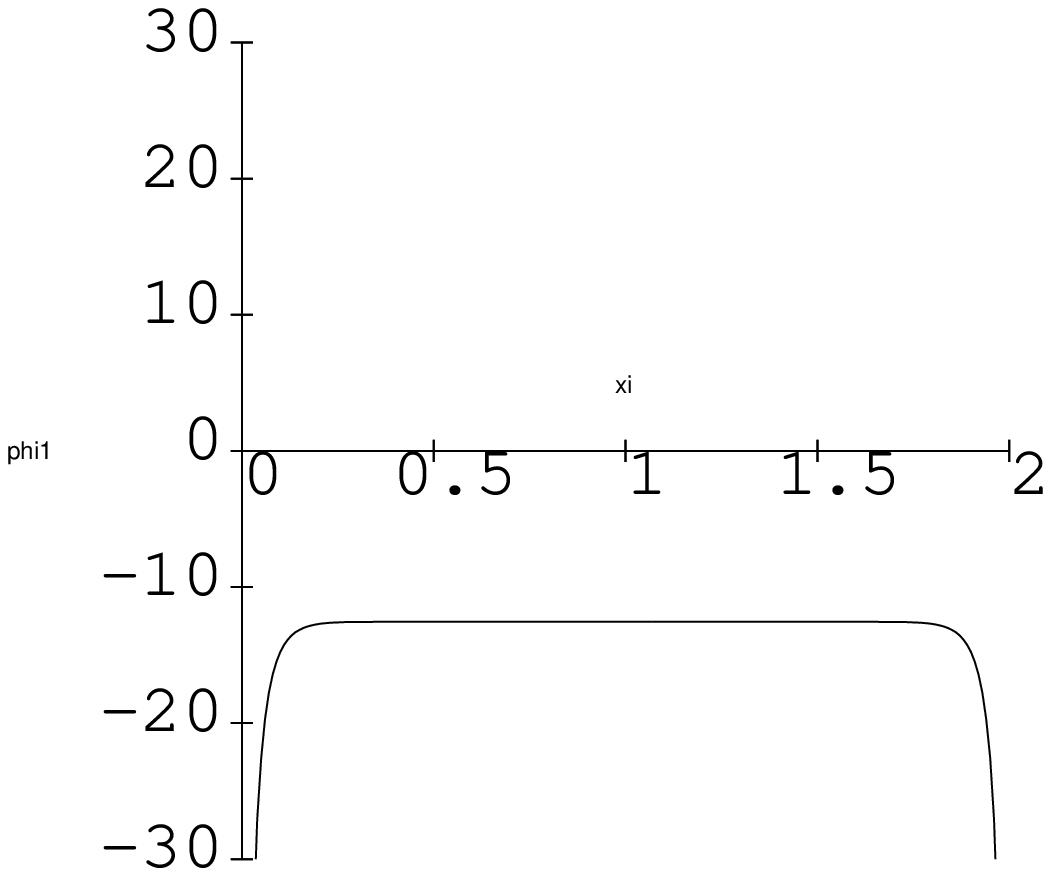,width=.32\textwidth}}
      \subfigure[$\varphi_2$]{\epsfig{figure=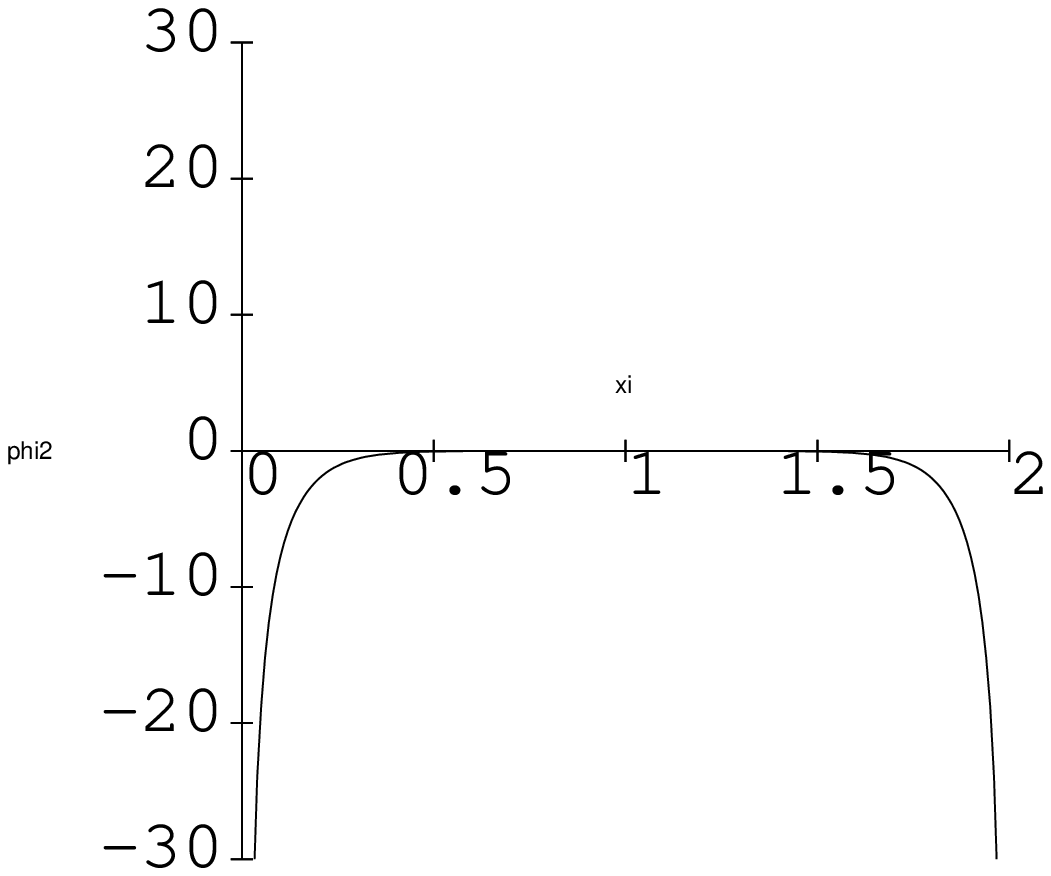,width=.32\textwidth}}
      \subfigure[$\varphi_3$]{\epsfig{figure=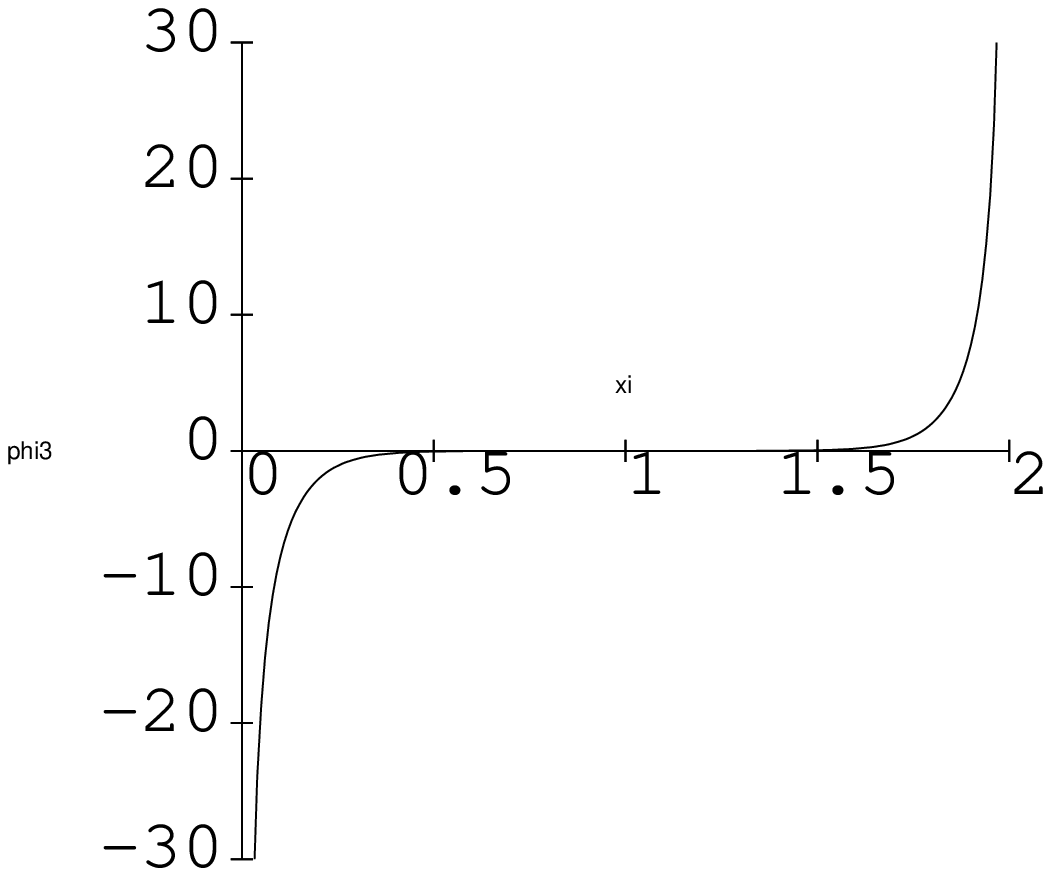,width=.32\textwidth}}}
\caption{Plots of the numerical solutions for the $\varphi_i$ at $t=0$ with $v_{init}=0.05$.}
\label{fig:varphi-t0}
\end{figure}

\begin{figure}
\centering
\psfrag{xi}{{\scriptsize$\xi$}}
\psfrag{phi1}{{\footnotesize$\varphi_1$}}
\psfrag{phi2}{{\footnotesize$\varphi_2$}}
\psfrag{phi3}{{\footnotesize$\varphi_3$}}
\mbox{
      \subfigure[$\varphi_1$]{\epsfig{figure=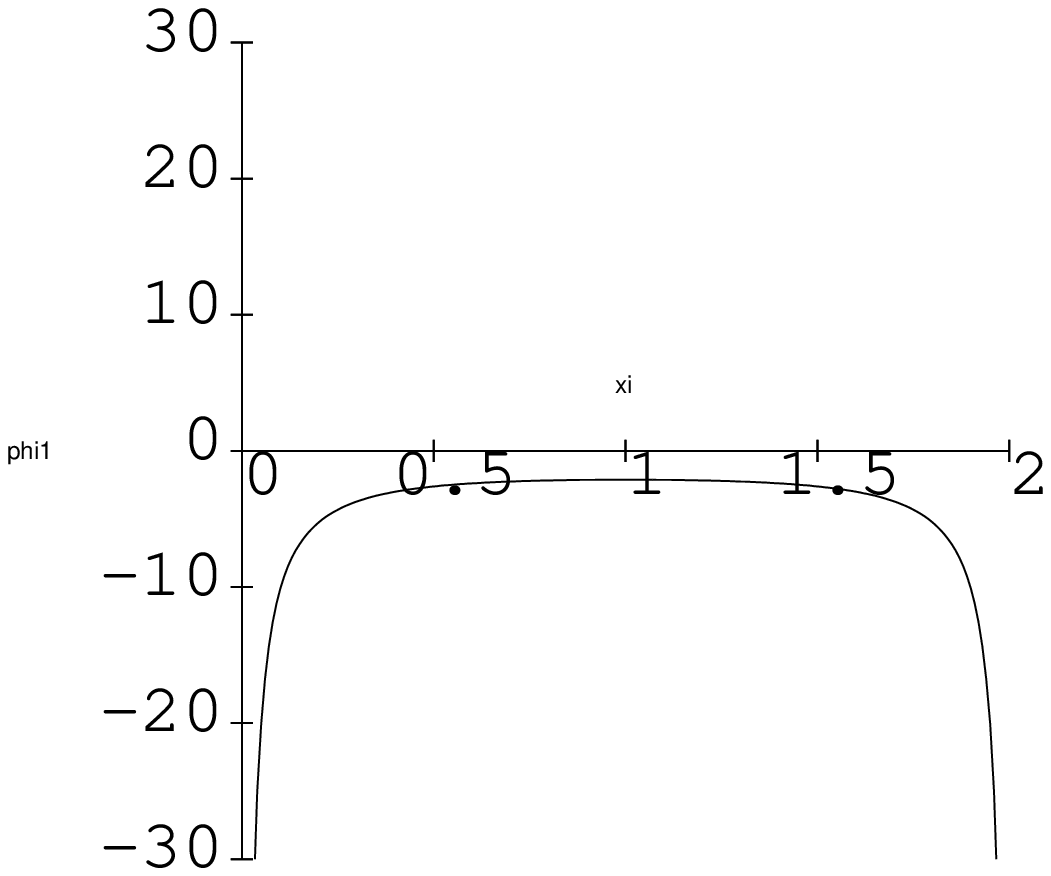,width=.32\textwidth}}
      \subfigure[$\varphi_2$]{\epsfig{figure=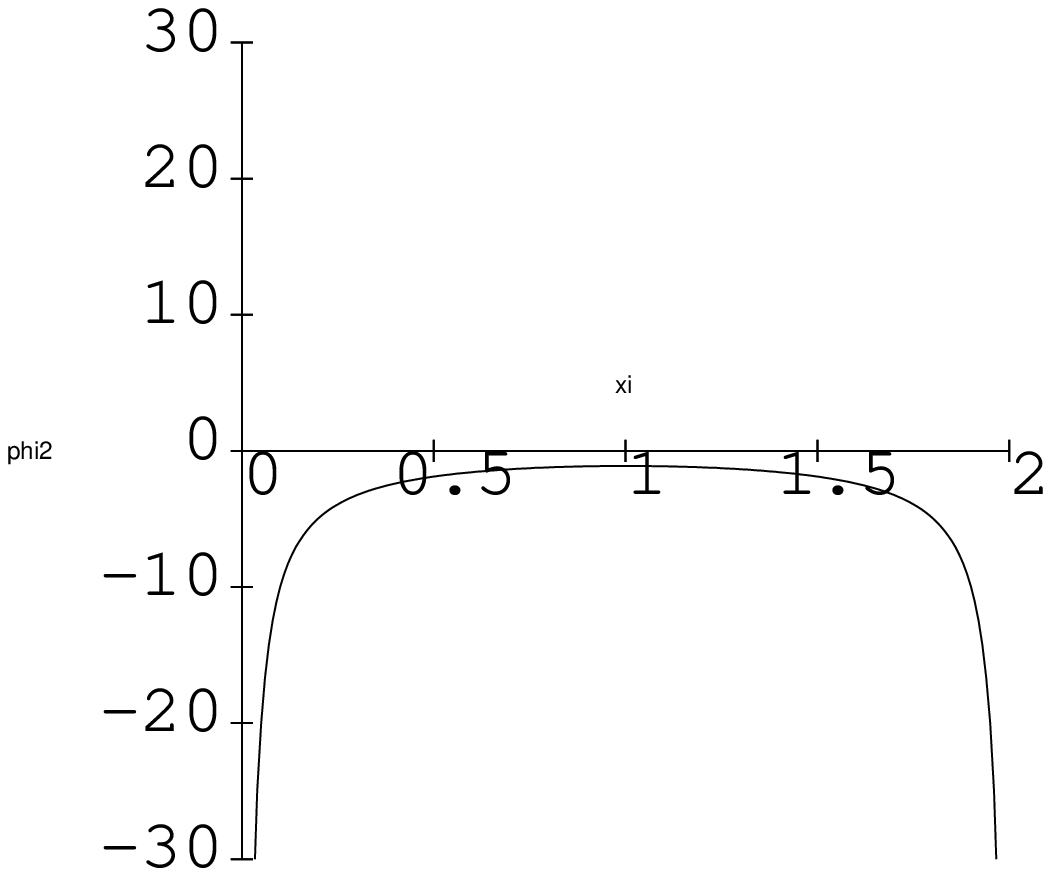,width=.32\textwidth}}
      \subfigure[$\varphi_3$]{\epsfig{figure=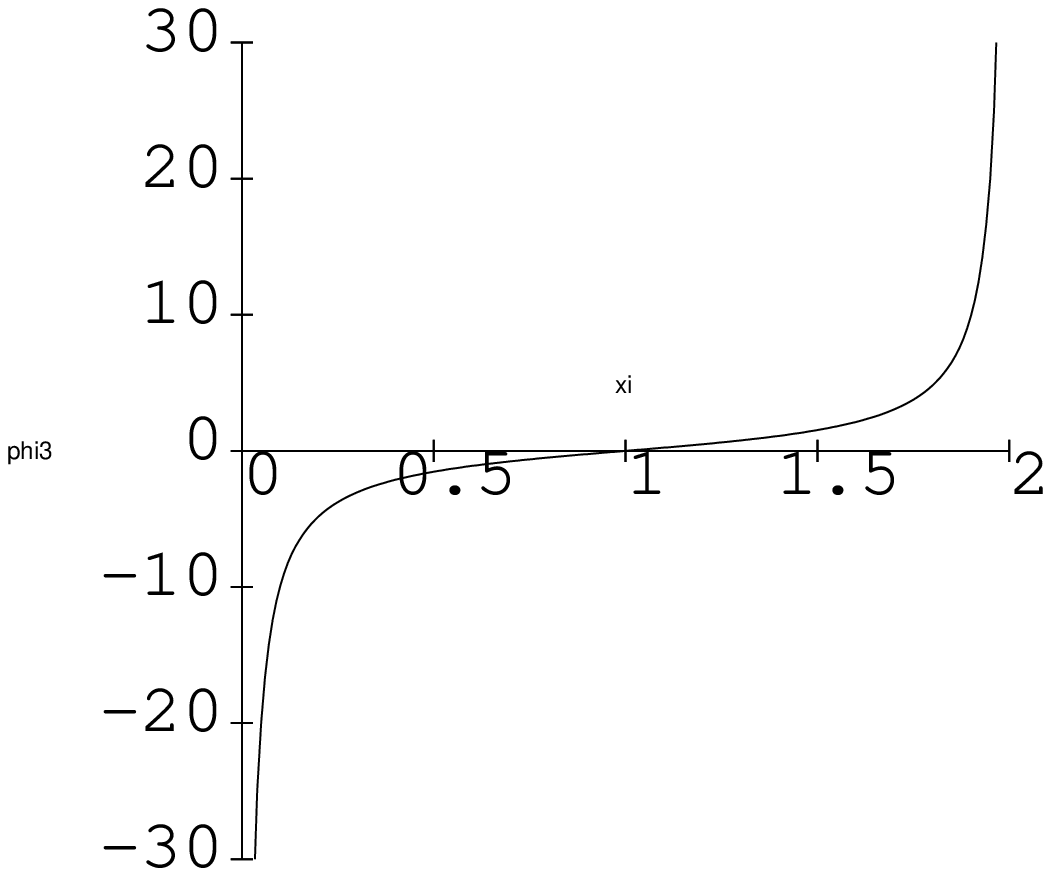,width=.32\textwidth}}}
\caption{Plots of the solutions for the $\varphi_i$ at $t=200$ with $v_{init}=0.05$.}
\label{fig:varphi-t200}
\end{figure}

\begin{figure}
\centering
\psfrag{xi}{{\scriptsize$\xi$}}
\psfrag{phi1}{{\footnotesize$\varphi_1$}}
\psfrag{phi2}{{\footnotesize$\varphi_2$}}
\psfrag{phi3}{{\footnotesize$\varphi_3$}}
\mbox{
      \subfigure[$\varphi_1$]{\epsfig{figure=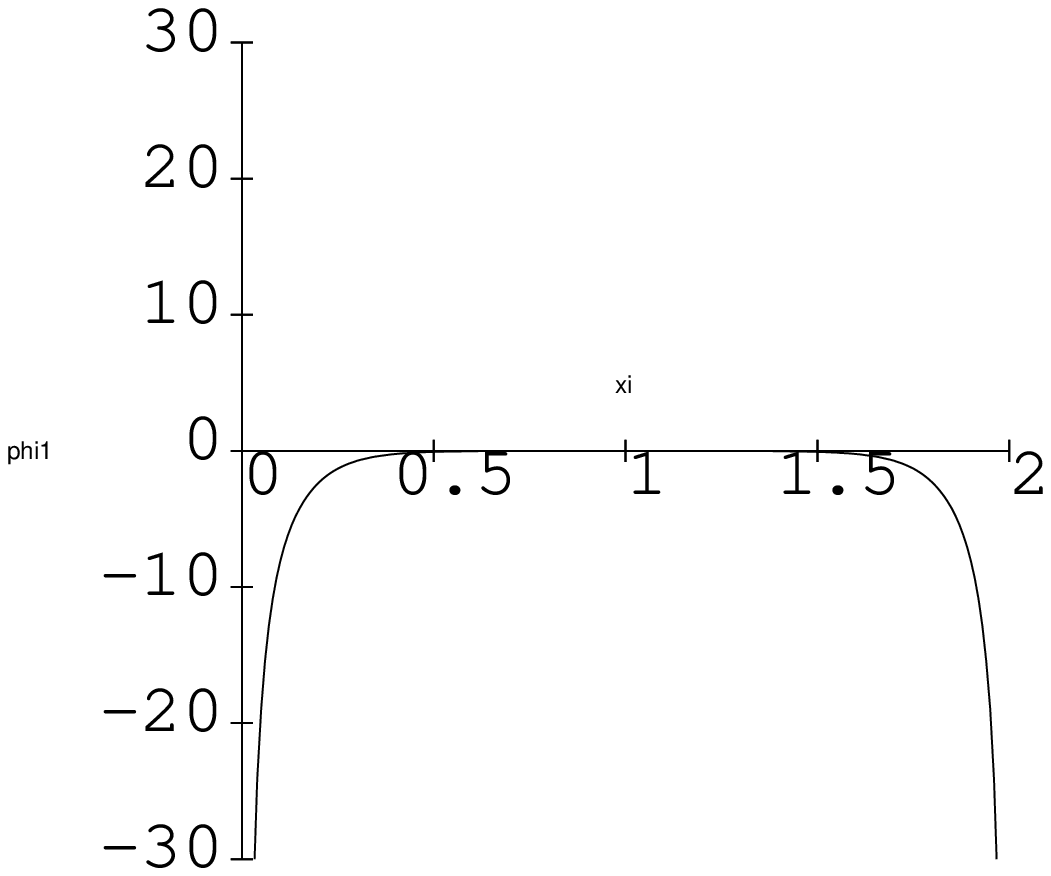,width=.32\textwidth}}
      \subfigure[$\varphi_2$]{\epsfig{figure=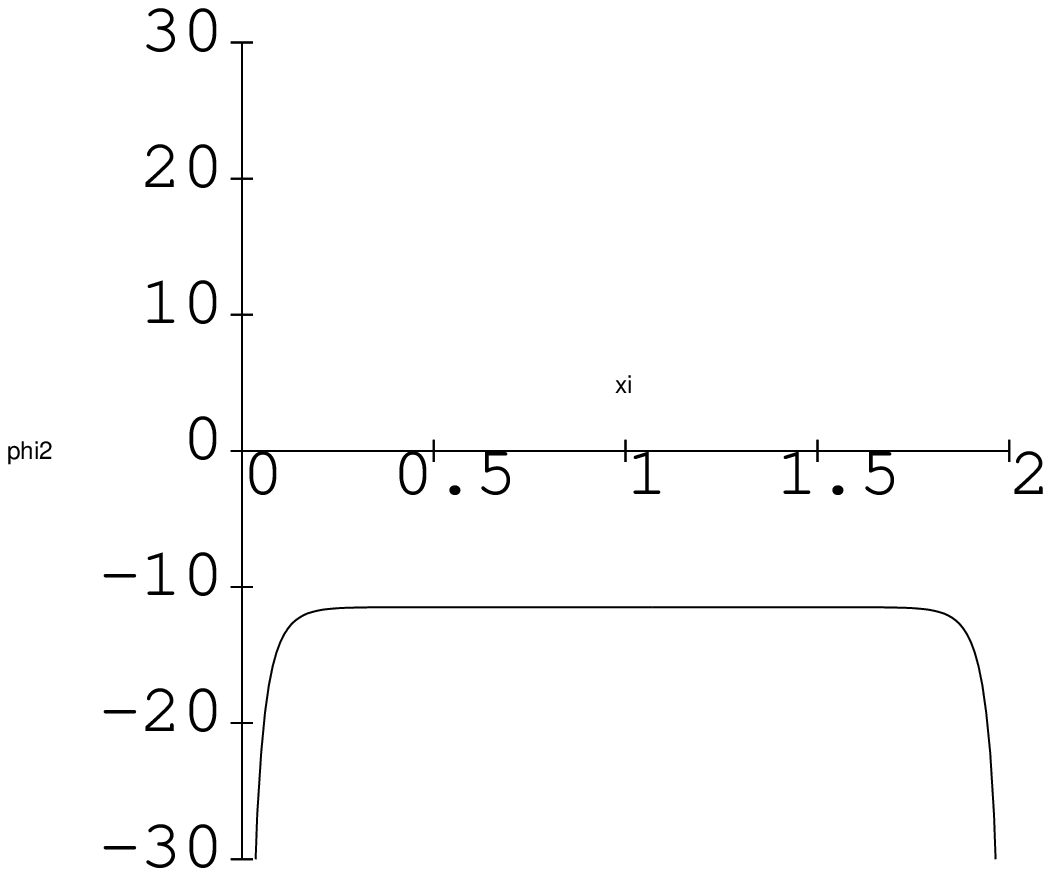,width=.32\textwidth}}
      \subfigure[$\varphi_3$]{\epsfig{figure=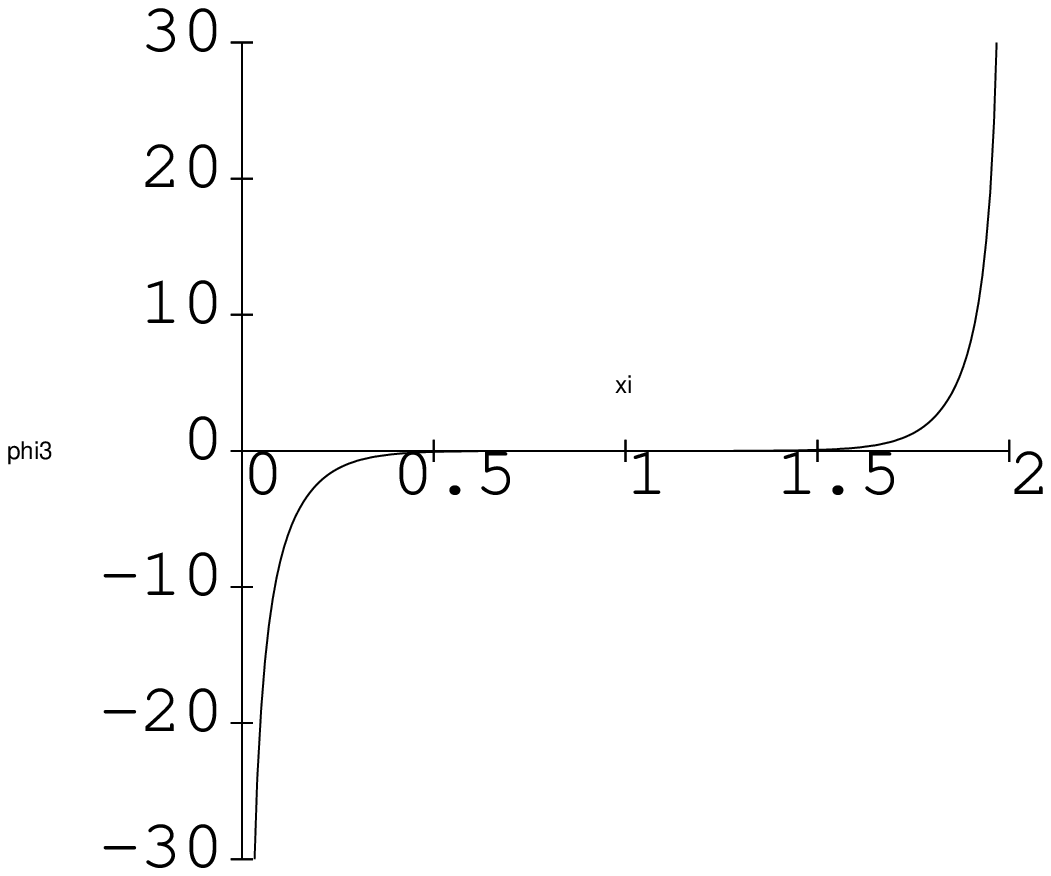,width=.32\textwidth}}}
\caption{Plots of the numerical solutions for the $\varphi_i$ at $t=400$ with $v_{init}=0.05$. }
\label{fig:varphi-t400}
\end{figure}

Figures \ref{fig:varphi-t0}, \ref{fig:varphi-t200} and \ref{fig:varphi-t400} show graphs of some of the solutions we obtained from our numerical program.

Figure \ref{fig:varphi-t0} shows the initial configuration for $\varphi_1$, $\varphi_2$ and $\varphi_3$. From these graphs we can see that $\varphi_1 \sim-K(0.9999999999) = -12.55264624$ and $\varphi_2\sim0$ and $\varphi_3\sim0$, except for the poles at $\xi=0$ and $\xi=2$, as we expected from equation \eqref{eq:f_kto1}.

Figure \ref{fig:varphi-t200} shows the solutions for $\varphi_1$, $\varphi_2$ and $\varphi_3$ close to the point of scattering. Here $\varphi_1 \approx \varphi_2$, which corresponds to the axisymmetric monopole solution (the `doughnut'). 

Figure \ref{fig:varphi-t400} shows the solutions for $\varphi_1$, $\varphi_2$ and $\varphi_3$ after scattering, with $t=400$. Note that after scattering $\varphi_1$ and $\varphi_2$ have exchanged roles, as expected. This corresponds to the D-strings scattering at $90^{\circ}$.

\section{Calculating the energy radiated}
\label{sec:energy-radiated}

In this section we describe the techniques we have used to calculate the energy radiated during scattering, using the numerical solutions for the $g_i$ from the previous section. 

The energy densities in terms of the $g_i$ are
\begin{eqnarray}
\label{eq:KEdensity}
\text{K.E. density} & = & \frac{T}{2} \left(\dot{g}_1^2 + \dot{g}_2^2 + \dot{g}_3^2\right)
\,\, , \\
\text{P.E. density} & = & \frac{T}{2} \bigg( g'^2_1 + g'^2_2 + g'^2_3
+  g_1^2g_2^2 + g_2^2g_3^2 + g_3^2g_1^2 
\nonumber \\
&& - \frac{2}{\xi} \left(g_1g_2(g_1 + g_2) + g_2g_3(g_2 + g_3) + g_3g_1(g_3 + g_1)\right)
\nonumber \\
&& + \frac{2}{\xi^2} (g_1^2 + g_2^2 + g_3^2 + g_1g_2 + g_2g_3 + g_3g_1)
\nonumber \\
&& - 2 (g'_1g_2g_3 + g'_2g_3g_1 + g'_3g_3g_1) 
\nonumber \\
\label{eq:PEdensity}
&& \left. + \frac{2}{\xi}(g_1(g'_2 + g'_3) + g_2(g'_3 + g'_1) + g_3(g'_1 + g'_2)) \right) \,\, .
\end{eqnarray}
Although there appear to be singularities in the potential energy density \eqref{eq:PEdensity} at $\xi=0$, all terms are in fact finite when we substitute in the series expansions for the $g_i$ \eqref{eq:g-series} (as was the case for the equations of motion \eqref{eq:g1eqn} - \eqref{eq:g3eqn}). We find
\begin{equation}
\text{P.E. density} (\xi = 0,t)  =  \frac{3T}{2} (a_1+b_1+c_1)^2 
 =  0 \,\, . \label{eq:lh-PEdensity}
\end{equation}
In our numerical programs to calculate the energy densities we used four $\xi$-points to calculate $g_i'$ in \eqref{eq:PEdensity} whenever possible, and we integrated the energy densities using Simpson's rule (see e.g. ref. \cite{Riley:1974}).

\subsection{Calculating the Energy in the $\varphi_i$}
\label{sec:varphi-energy}

In this section we present the results of our energy calculations for the $\varphi_i$ (i.e. the total energy in the zero modes and the non-zero modes). We will present all energies as ratios to the initial total energy, which we denote $E_{tot}^{init}$.

First we present the ratio of the total energy in the $\varphi_i$, $E_{tot}$, to $E_{tot}^{init}$ at different times. The results are given in table \ref{table:table1} for $v_{init}=0.05$ and in table \ref{table:table2} for $v_{init}=0.1$. If there were no numerical inaccuracies in our results this ratio would be 1 at all times because the total energy is conserved (see section \ref{sec:energy-cons}). Therefore the order of magnitude at which $E_{tot}/E_{tot}^{init}$ deviates from 1 at time $t$ gives us an estimate of the numerical inaccuracy in our calculation at that time.

\begin{table}[p]
\centering
\begin{tabular}{cc}
\begin{tabular}{|c|c|}
\hline 
 Time & $E_{tot}/E^{init}_{tot}$ 
\\\hline
0  &  1 
\\\hline
25    &    0.99999999831 
\\\hline
50    &    0.99999999658 
\\\hline
75    &    0.99999999639 
\\\hline
100   &    0.99999999681 
\\\hline
125   &    0.99999999762 
\\\hline
150   &    0.99999999706 
\\\hline
175   &    0.99999999673 
\\\hline
200   &    0.99999999698 
\\\hline
225   &    0.99999999709 
\\\hline
250   &    0.99999999718 
\\\hline
275   &    0.99999999683 
\\\hline
300   &    0.9999999968 
\\\hline
325   &    0.99999999736 
\\\hline
350   &    0.99999999478 
\\\hline
375   &    0.99999999561 
\\\hline
\end{tabular}
\begin{tabular}{|c|c|}
\hline 
 Time & $E_{tot}/E^{init}_{tot}$
\\\hline
400   &    0.99999999607 
\\\hline
425   &    0.99999999663
\\\hline 
450   &    0.99999999671 
\\\hline
475   &    0.99999999891 
\\\hline
500   &    0.99999999925 
\\\hline
525   &    0.9999999971 
\\\hline
550   &    0.99999999391 
\\\hline
575   &    0.99999999648 
\\\hline
600   &    1.0000000008 
\\\hline
625   &    0.99999999938 
\\\hline
650   &    1.0000000012 
\\\hline
675   &    0.99999999632 
\\\hline
700   &    0.99999999628 
\\\hline
725   &    1.0000000017 
\\\hline
750   &    1.0000000045 
\\\hline
775   &    1.0000000081 
\\\hline
\end{tabular}
\end{tabular}
\caption{Table showing the ratio of the total energy $E_{tot}$ to the total initial energy $E_{tot}^{init}$ at different times, with initial velocity $v_{init} = 0.05$.}
\label{table:table1}
\end{table}


\begin{table}[p]
\centering
\begin{tabular}{cc}
\begin{tabular}{|c|c|}
\hline  
Time & $E_{tot}/E^{init}_{tot}$ 
\\\hline
0     &         1 
\\\hline
10    &   0.99999999945 
\\\hline
20    &   0.99999999947 
\\\hline
30    &   0.99999999974 
\\\hline
40    &   1.0000000001 
\\\hline
50    &   0.99999999987 
\\\hline
60    &   0.99999999995 
\\\hline
70    &   1.0000000004 
\\\hline
80    &   1.0000000003 
\\\hline
90    &   1.0000000004 
\\\hline
100   &   1.0000000004 
\\\hline
110   &   1.0000000004 
\\\hline
120   &   1.0000000004 
\\\hline
130   &   1.0000000005 
\\\hline
140   &   1.0000000006 
\\\hline
150   &   1.0000000003 
\\\hline
\end{tabular} &
\begin{tabular}{|c|c|}
\hline Time & $E_{tot}/E^{init}_{tot}$ 
\\\hline
160    &    1.0000000001
\\\hline 
170    &    1.0000000007 
\\\hline
180    &    1.0000000008 
\\\hline
190    &    1.0000000009 
\\\hline
200    &    1.0000000008 
\\\hline
210    &    1.0000000014 
\\\hline
220    &    1.0000000011 
\\\hline
230    &    1.0000000007 
\\\hline
240    &    1.0000000001 
\\\hline
250    &    1.0000000005 
\\\hline
260    &    0.99999999947 
\\\hline
270    &    0.99999999825 
\\\hline
280    &    0.99999999828 
\\\hline
290    &    0.99999999844 
\\\hline
300    &    0.9999999983 
\\\hline
310    &    0.99999999936 
\\\hline
\end{tabular}
\end{tabular}
\caption{Table showing the ratio of the total energy $E_{tot}$ to the total initial energy $E_{tot}^{init}$ at different times, with initial velocity $v_{init} = 0.1$.}
\label{table:table2}
\end{table} 

\begin{itemize}
\item[a)]
From table \ref{table:table1}, for $v_{init}=0.05$, the numerical inaccuracy in the total energy is around $10^{-8}$.
\item[b)]
From table \ref{table:table2}, for $v_{init}=0.1$, the numerical inaccuracy in the total energy is around $10^{-9}$.
\end{itemize}

\begin{figure}
\centering
\psfrag{PE}[r][l]{PE}
\psfrag{time}{$t$}
\psfrag{1e-12}{{\footnotesize$10^{-12}$}}
\psfrag{1e-11}{{\footnotesize$10^{-11}$}}
\psfrag{1e-10}{{\footnotesize$10^{-10}$}}
\psfrag{1e-09}{{\footnotesize$10^{-9}$}}
\psfrag{1e-08}{{\footnotesize$10^{-8}$}}
\psfrag{1e-07}{{\footnotesize$10^{-7}$}}
\psfrag{1e-06}{{\footnotesize$10^{-6}$}}
\psfrag{1e-05}{{\footnotesize$10^{-5}$}}
\psfrag{.1e-3}{{\footnotesize$10^{-4}$}}
\psfrag{300}{{\footnotesize300}}
\psfrag{250}{{\footnotesize250}}
\psfrag{200}{{\footnotesize200}}
\psfrag{150}{{\footnotesize150}}
\psfrag{100}{{\footnotesize100}}
\psfrag{50}{{\footnotesize50}}
\psfrag{0}{{\footnotesize0}}
\psfrag{400}{{\footnotesize400}}
\psfrag{600}{{\footnotesize600}}
\mbox{\subfigure[$v_{init} = 0.05$]{\epsfig{figure=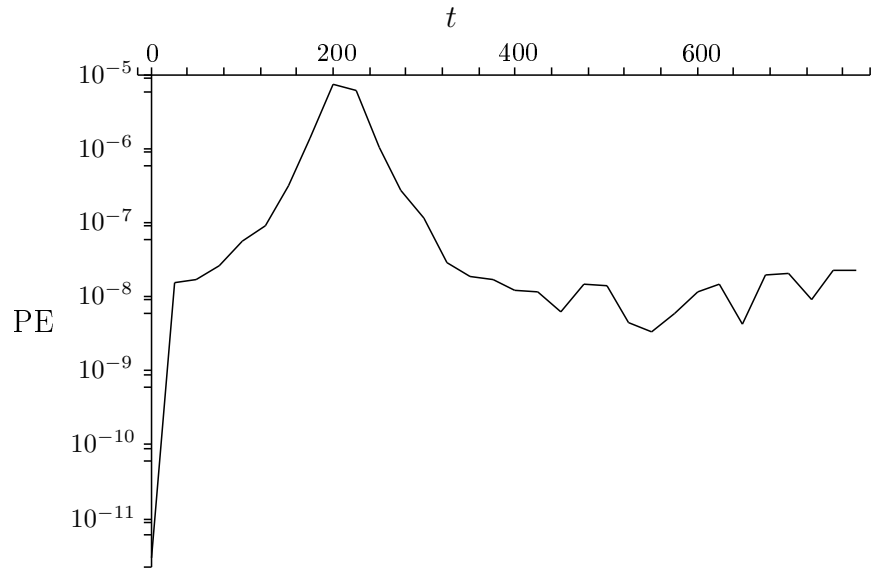,width=.80\textwidth}\label{subfigure:varphi-PE-1}}
     } 
\mbox{\subfigure[$v_{init} = 0.1$]{\epsfig{figure=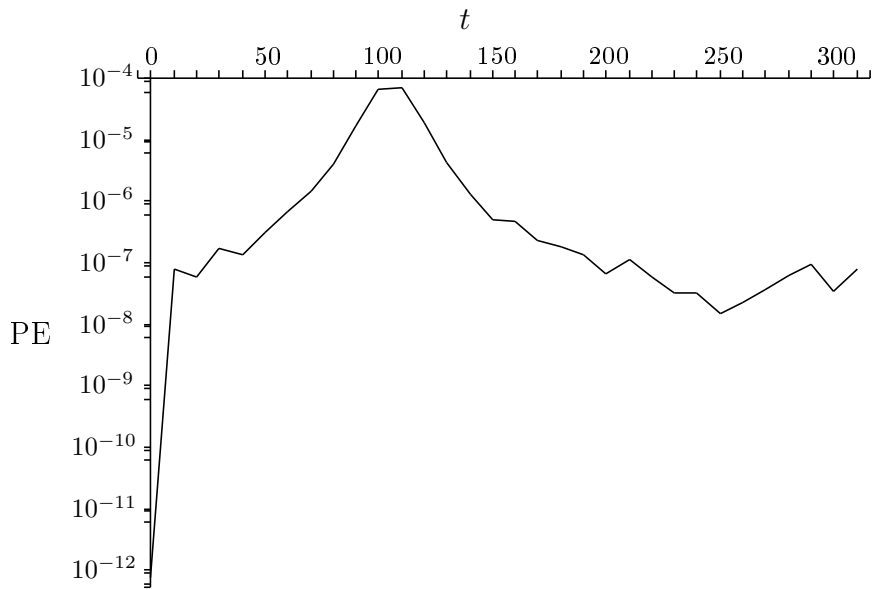,width=.80\textwidth}\label{subfigure:varphi-PE-2}}
     } 
\caption{Logarithmic plot of the potential energy in the $\varphi_i$ against time}
\end{figure}

Next we present the potential energy in the $\varphi_i$ in figure \ref{subfigure:varphi-PE-1} for $v_{init}=0.05$, and in figure \ref{subfigure:varphi-PE-2} for $v_{init}=0.1$. As we pointed out in section \ref{sec:energy-densities}, the potential energy measures the deviation of the solution from the solutions to the BPS equations, the $f_i$. So the potential energy originates entirely from the non-zero modes. In section \ref{sec:decoupling_proof} we found that the non-zero modes behave like harmonic oscillators when the D-strings are far apart. So at late times their kinetic energy is of the same order as their potential energy, and so the magnitude of the potential energy in the $\varphi_i$ is approximately half  the total energy in the non-zero modes. 
\begin{itemize}
\item[a)]
In the graph in figure \ref{subfigure:varphi-PE-1}, for $v_{init}=0.05$, we can see that the potential energy increases up to around $10^{-5}$ near the point of scattering $t\approx 200$. After scattering the potential energy decreases back down to order $10^{-8}$, which is the order of the numerical inaccuracies in this calculation. This suggests that all the energy has been transferred back into the zero modes after scattering, and therefore no energy has been radiated.
\item[b)]
Similarly for $v_{init}=0.1$, in the graph in figure \ref{subfigure:varphi-PE-2}, we find the potential energy increases up to the order of $10^{-4}$ around the point of scattering at $t \approx 100$. Then it decreases back down to the order of $10^{-7}$ after scattering. Although this is slightly higher than the order of numerical inaccuracy, it is still much lower than we would expect from the prediction \eqref{eq:Erad}, which would give $E_{rad}/E_{tot} \sim 10^{-3}$.
\end{itemize}   

\subsection{Calculating the energy in the non-zero modes directly}
\label{sec:epsilon-energy}

In the previous section we deduced the energy radiated from the potential energy of the full numerical solutions for the $g_i$. In the next section we will calculate the energy in the non-zero modes directly. In order to do this we need to separate out the non-zero modes $\epsilon_i$ from the full solutions $\varphi_i$, where
\begin{equation}
\varphi_i(\xi,t) = f_i(\xi,k(t)) + \epsilon_i(\xi,t) \,\, .
\end{equation}
We will work in the asymptotic limit, when the D-strings are far apart, and we can use $K(k(t))$ as the modular parameter instead of $k(t)$ (see section \ref{sec:zero_mode}). We can also use the approximations \eqref{eq:f_kto1}.

In order to separate out the zero modes from the non-zero modes it is necessary to calculate $K$ and $\dot{K}$ for a given numerical solution for the $\varphi_i$. A first approximation for $K$, call it $\tilde{K}$, is $\tilde{K}=-\varphi_2(\xi=1,t)$. This assumes that $\epsilon_2(\xi=1,t)=0$. To obtain a more accurate approximation for $K$, call it $\hat{K}$, we can use the fact that the non-zero modes are harmonic oscillators in the asymptotic limit (see section \ref{sec:decoupling_proof}). We set
\begin{equation}
\varphi_2(1,t) = -\hat{K} + \delta \,\, ,
\end{equation}
where $\delta=\epsilon_2(1,t)$ is chosen such that the integral $\int_0^2 \epsilon_2(\xi,t)d\xi=0$. We then have
\begin{equation}
\epsilon_i(\xi,t) = \varphi_i(\xi,t) - f_i(\xi,\hat{K}) \,\, .
\end{equation}
We can calculate an approximation for $\dot{K}$ and the $\dot{\epsilon}_i$ using a similar procedure to that described above.

In order to calculate the energy in the $\epsilon_i$ we have to assume that the zero modes and non-zero modes have decoupled from one another, as is the case when the D-strings are far apart. Then the kinetic energy density and potential energy density for the $\epsilon_i$ are given by
\begin{eqnarray}
\label{eq:KEdensity-epsilon}
\text{K.E. density} & = & \frac{T}{2} (\dot{\epsilon}_1^2 + \dot{\epsilon}_2^2 + \dot{\epsilon}_3^2) \,\, ,
\\
\text{P.E. density} & = & T \Big( \frac{1}{2} (\epsilon'^2_1 + \epsilon'^2_2 + \epsilon'^2_3)
\nonumber \\
&& + \frac{1}{\xi^2} (\epsilon_1^2 + \epsilon_2^2 + \epsilon_3^2 + \epsilon_1\epsilon_2 + \epsilon_2\epsilon_3 + \epsilon_3\epsilon_1)
\nonumber \\
\label{eq:PEdensity-epsilon}
&& + \frac{1}{\xi}(\epsilon_1(\epsilon'_2 + \epsilon'_3) + \epsilon_2(\epsilon'_3 + \epsilon'_1) + \epsilon_3(\epsilon'_1 + \epsilon'_2)) \Big)\,\, ,
\end{eqnarray}
where we have neglected all terms of order $\epsilon^3$ and higher in the potential energy density \eqref{eq:PEdensity-epsilon}.

\begin{figure}
\centering
\psfrag{time}{$t$}
\psfrag{KE}{KE}
\psfrag{PE}{PE}
\psfrag{1e-08}{{\scriptsize$10^{-8}$}}
\psfrag{1e-07}{{\scriptsize$10^{-7}$}}
\psfrag{1e-06}{{\scriptsize$10^{-6}$}}
\psfrag{1e-05}{{\scriptsize$10^{-5}$}}
\psfrag{.1e-3}{{\scriptsize$10^{-4}$}}
\psfrag{.1e-2}{{\scriptsize$10^{-3}$}}
\psfrag{.1e-1}{{\scriptsize$10^{-2}$}}
\psfrag{.1}{{\scriptsize$10^{-1}$}}
\psfrag{5e-08}{{\scriptsize$5 \times 10^{-8}$}}
\psfrag{5e-09}{{\scriptsize$5 \times 10^{-9}$}}
\psfrag{400}{{\scriptsize400}}
\psfrag{500}{{\scriptsize500}}
\psfrag{600}{{\scriptsize600}}
\psfrag{700}{{\scriptsize700}}
\mbox{\subfigure[kinetic energy]{\epsfig{figure=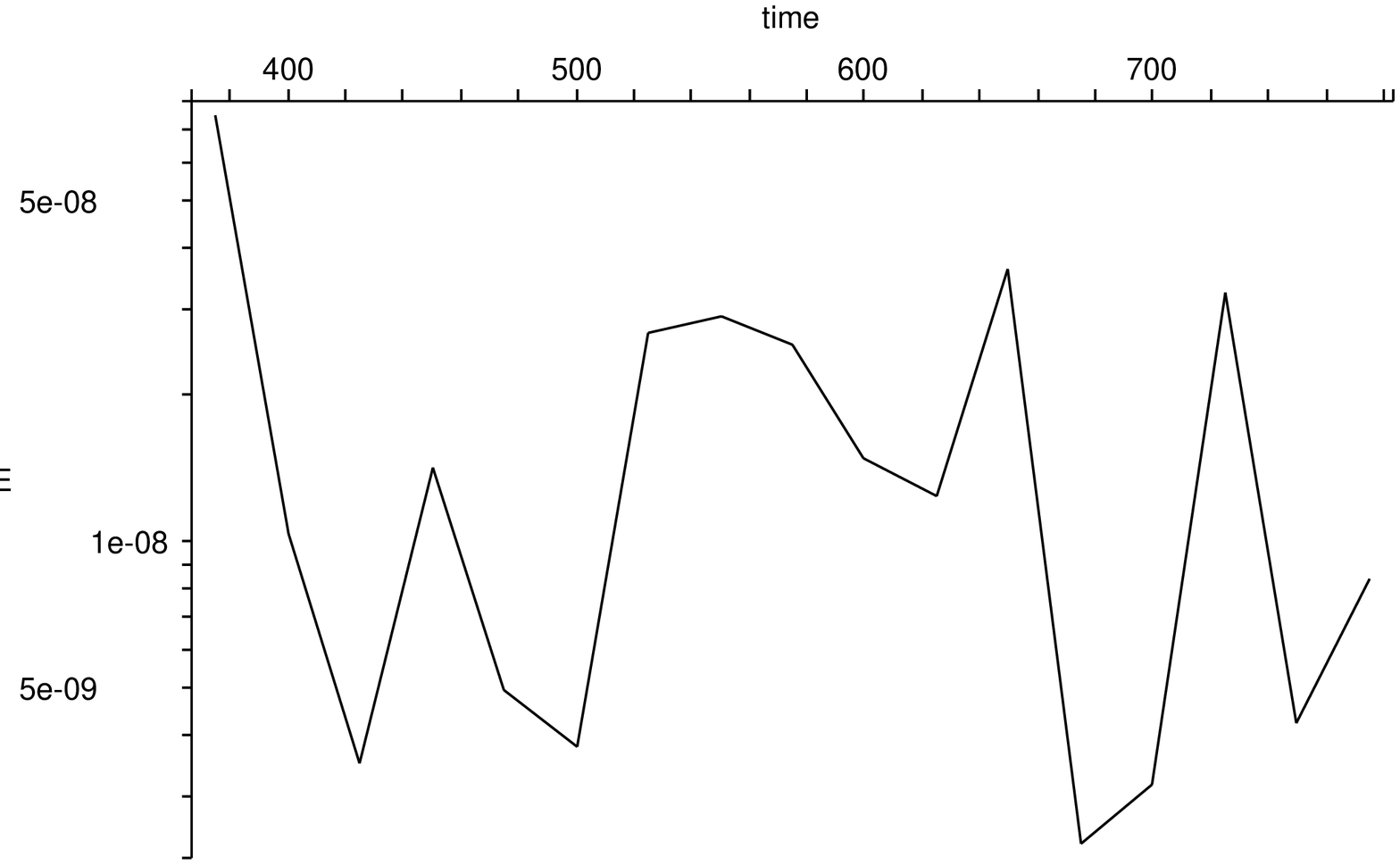,width=.7\textwidth}}
     }
\mbox{\subfigure[potential energy]{\epsfig{figure=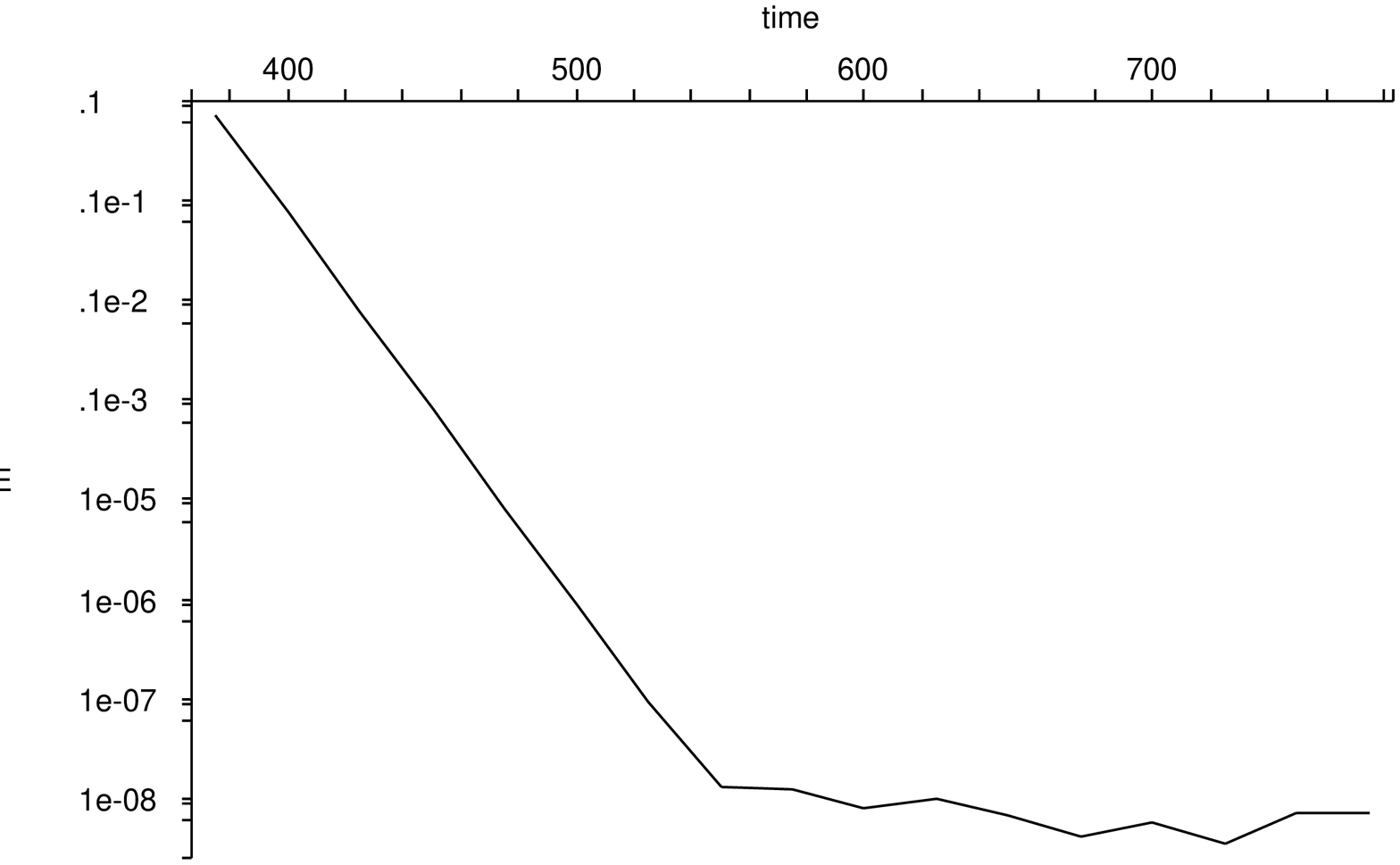,width=.7\textwidth}}
     }
\caption{Graphs showing the kinetic and potential energy densities in the $\epsilon_i$ for $v_{init}=0.05$.}
\label{fig:epsilon-energy-1}
\end{figure}

\begin{figure}
\centering
\psfrag{time}{$t$}
\psfrag{KE}{KE}
\psfrag{PE}{PE}
\psfrag{1e-07}{{\scriptsize$10^{-7}$}}
\psfrag{1e-06}{{\scriptsize$10^{-6}$}}
\psfrag{1e-05}{{\scriptsize$10^{-5}$}}
\psfrag{.1e-3}{{\scriptsize$10^{-4}$}}
\psfrag{.1e-2}{{\scriptsize$10^{-3}$}}
\psfrag{8e-08}{{\scriptsize$8 \times 10^{-8}$}}
\psfrag{6e-08}{{\scriptsize$6 \times 10^{-8}$}}
\psfrag{4e-08}{{\scriptsize$4 \times 10^{-8}$}}
\psfrag{200}{{\scriptsize200}}
\psfrag{220}{{\scriptsize220}}
\psfrag{240}{{\scriptsize240}}
\psfrag{260}{{\scriptsize260}}
\psfrag{280}{{\scriptsize280}}
\psfrag{300}{{\scriptsize300}}
\mbox{\subfigure[kinetic energy]{\epsfig{figure=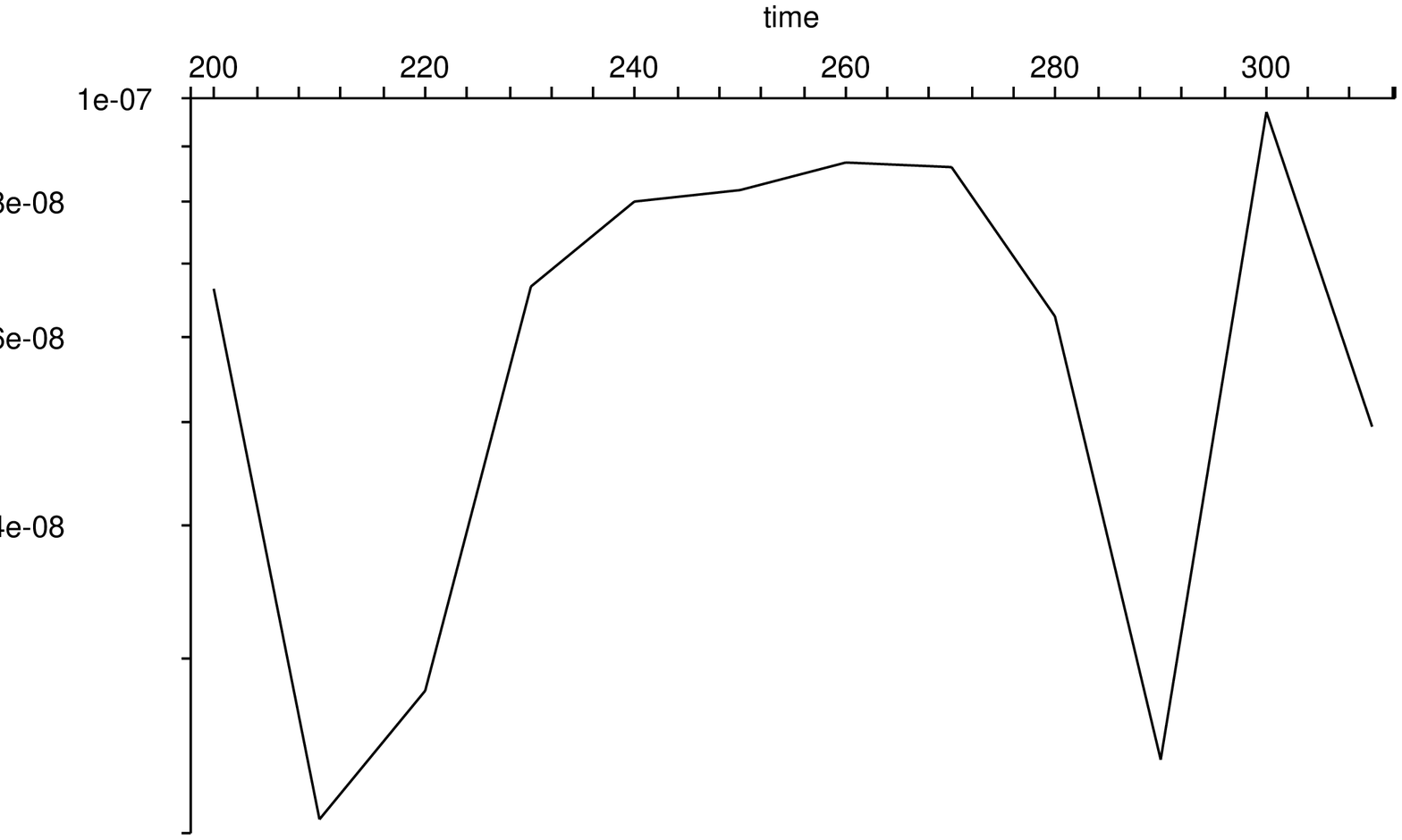,width=.6\textwidth}}
      }
\mbox{\subfigure[potential energy]{\epsfig{figure=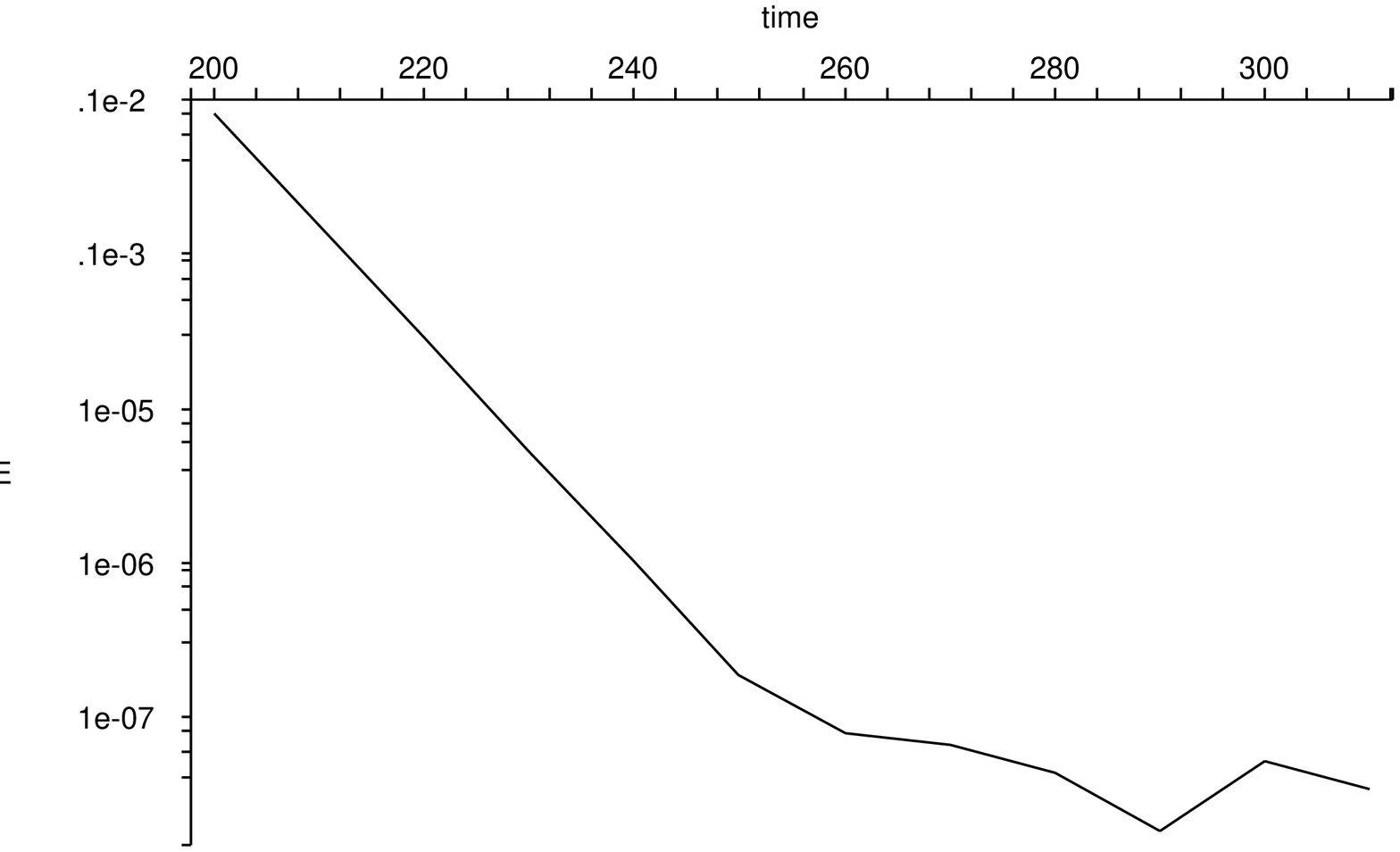,width=.6\textwidth}}
     }
\caption{Graphs showing the kinetic and potential energy densities in the $\epsilon_i$ for $v_{init}=0.1$.}
\label{fig:epsilon-energy-2}
\end{figure}

The graphs in figures \ref{fig:epsilon-energy-1} and \ref{fig:epsilon-energy-2} show the potential and kinetic energies calculated for $v_{init}=0.05$ and $v_{init}=0.1$ respectively. 
\begin{itemize}
\item[a)]
In figure \ref{fig:epsilon-energy-1}, for $v_{init}=0.05$, the potential energies are much higher than those found in section \ref{sec:varphi-energy} for $400 < t < 525$ because our calculations were only valid in the asymptotic limit. For $t \geq 550$ the potential energies are of the order $10^{-8}$, which agrees with the results presented in section \ref{sec:varphi-energy}. The $\epsilon_i$ we have calculated could have originated entirely from numerical errors.
\item[b)]
Similarly for $v_{init}=0.1$, for later times $t \geq 260$, the potential energies are of the order $10^{-7}$, which agrees with the results presented in section \ref{sec:varphi-energy}. 
\end{itemize}

\section{Discussion}
\label{sec:conclusions}

We have solved numerically the equations of motion derived from the Yang-Mills limit of the Born-Infeld action for D-strings, starting with two D-strings stretched between two D3-branes a long way apart from each other in the D3-brane worldvolume, and moving towards each other very slowly. We calculated numerically the energy radiated during the scattering of the two D-strings.

In discussing our results we should bear in mind that the Born-Infeld action is only approximate, and is inaccurate in regions of space which are highly curved. We discussed this issue in ref. \cite{Barrett:2004zt}, and gave arguments as to why our results may still be relevant. In short, there is evidence to suggest that the solutions to Nahm's equations \eqref{eq:nahm} are solutions of the full string theory (see ref. \cite{Hashimoto:2005yy} for a recent discussion\footnote{We thank A.Gustavsson for bringing this paper to our attention.}). The configurations we used were close to solutions of Nahm's equations, as is demonstrated by equation \eqref{eq:varphi}. It is then reasonable to assume that the motion is accurately described by the Born-Infeld action. 

Our numerical results reproduce the $90^{\circ}$ scattering which we expected from the comparison to monopole scattering. It is encouraging that this follows directly from the equations of motion, rather than having to be put in by hand, as we did in ref. \cite{Barrett:2004zt}.

The prediction of ref. \cite{Manton:1988bn} that $E_{rad}/E_{tot} \sim 1.35 v_{-\infty}^3$ gives
\begin{equation}
\frac{E_{rad}}{E_{tot}} \sim 10^{-4} \quad \textrm{for $v_{-\infty}=0.05$} \,\, , \quad \frac{E_{rad}}{E_{tot}} \sim 10^{-3} \quad \textrm{for $v_{-\infty} = 0.1$ } \,\, .
\end{equation}
(Note that the initial values for our program were $v_{init}=0.05$ and $v_{init}=0.1$, rather than $v_{-\infty}=0.05$ and $v_{-\infty}=0.1$. However, since the D-strings were very far apart initially, it is safe to take $v_{-\infty}=0.05$ and $v_{-\infty}=0.1$ for our results). Our results for the energy radiated are 
\begin{equation}
\frac{E_{rad}}{E_{tot}^{init}} \sim 10^{-8} \quad \textrm{for $v_{-\infty}=0.05$} \,\, , \quad \frac{E_{rad}}{E_{tot}^{init}} \sim  10^{-7} \quad \textrm{for $v_{-\infty} = 0.1$ } \,\, .
\label{eq:numerical_result}
\end{equation}
(See figures \ref{subfigure:varphi-PE-1} and \ref{fig:epsilon-energy-1} for $v_{-\infty}=0.05$ and \ref{subfigure:varphi-PE-2} and \ref{fig:epsilon-energy-2} for $v_{-\infty}=0.1$). Our results therefore suggest that the energy radiated is much smaller then the prediction of ref. \cite{Manton:1988bn} suggests. Indeed, our values for the energy radiated in \eqref{eq:numerical_result} are approximately of the same order as the numerical inaccuracy in our programs (see the discussion surrounding tables \ref{table:table1} and \ref{table:table2}). So our results are consistent with there being no energy radiated during D-string scattering. It would be nice to be able to support this conclusion with further evidence from the Born-Infeld action.

\section*{Acknowledgements}

This work was partly supported by the EC network ``EUCLID", contract number HPRN-CT-2002-00325. JKB was supported by an EPSRC studentship. We thank Ed Corrigan, Simon Ross, Douglas Smith and Clifford Johnson for useful discussions.

\appendix

\section{Expansions of $f_1$, $f_2$ and $f_3$ in the Limit $k\to 1$}

In this appendix we derive the series expansions of $f_1$, $f_2$ and $f_3$ in the limit $k \to 1$ that we used in section \ref{sec:asymptotic_calculations}. 

\subsection{Expanding in $k'(k)$}
\label{sec:k-series}

In order to obtain $f_1$, $f_2$ and $f_3$ as functions of $k'$ we will use the following transformation of elliptic functions, which can be found in ref. \cite{Erdelyi}. Under the transformation
\begin{equation}
\xi  \to  \tilde{\xi} = -i\xi \,\, , \quad
k  \to  \tilde{k} = k' = \sqrt{1-k^2} \,\, ,
\end{equation}
the elliptic functions transform as
\begin{eqnarray}
K(k) & \to & \tilde{K}(k) = K'(k) = K(k')\,\, , \nonumber \\
\rm{sn}(\xi,k) & \to & \rm{sn}(\tilde{\xi},\tilde{k}) = -i\frac{\rm{sn}(\xi,k)}{\rm{cn}(\xi,k)} \,\, , \quad
\rm{cn}(\xi,k)  \to  \rm{cn}(\tilde{\xi},\tilde{k}) = \frac{1}{\rm{cn}(\xi,k)} \,\, , \nonumber \\
\rm{dn}(\xi,k) & \to & \rm{dn}(\tilde{\xi},\tilde{k}) = \frac{\rm{dn}(\xi,k)}{\rm{cn}(\xi,k)} \,\, . 
\label{eq:f_trans}
\end{eqnarray}
Substituting \eqref{eq:f_trans} into the expressions \eqref{eq:fsoln} for $f_1$, $f_2$ and $f_3$ we obtain
\begin{eqnarray}
f_1(\xi,k) & = & -iK(k) \frac{\rm{cn}(iK(k)\xi,k')}{\rm{sn}(iK(k)\xi,k')} \,\, , \quad
f_2(\xi,k)  =  -iK(k) \frac{\rm{dn}(iK(k)\xi,k')}{\rm{sn}(iK(k)\xi,k')} \,\, , \nonumber \\
f_3(\xi,k) & = & -iK(k) \frac{1}{\rm{sn}(iK(k)\xi,k')} \,\, .
\label{eq:f_transformed}
\end{eqnarray}
We will also use from ref. \cite{Erdelyi} the expansions for $\rm{sn}(\xi,k)$, $\rm{cn}(\xi,k)$ and $\rm{dn}(\xi,k)$ for small $k$. These are
\begin{align}
\rm{sn}(\xi,k) &  = & \!\!\!\!\!\! \frac{2\pi}{K(k)}  \bigg( & \frac{1}{4} \sin\left(\frac{\pi\xi}{2K(k)}\right) + \frac{5k^2}{64} \sin\left(\frac{\pi\xi}{2K(k)}\right) + \frac{k^2}{64} \sin\left(\frac{3\pi\xi}{2K(k)}\right) 
\nonumber \\
&&& + \cdots \bigg) \,\, , \nonumber \\
\rm{cn}(\xi,k)&  = & \!\!\!\!\!\!\frac{2\pi}{K(k)} \bigg(& \frac{1}{4} \cos\left(\frac{\pi\xi}{2K(k)}\right) + \frac{3k^2}{64} \cos\left(\frac{\pi\xi}{2K(k)}\right) + \frac{k^2}{64} \sin\left(\frac{3\pi\xi}{2K(k)}\right) \nonumber \\
&&& + \cdots \bigg) \,\, , \nonumber \\
\rm{dn}(\xi,k)&  = & \!\!\!\!\!\!\frac{2\pi}{K(k)} \bigg( & 1 + \frac{k^2}{4} \cos\left(\frac{\pi\xi}{2K(k)}\right) +  \cdots \bigg) \,\, . 
\end{align}
Substituting these into the expressions \eqref{eq:f_transformed} for $f_1$, $f_2$ and $f_3$ we find
\begin{align}
f_1(\xi,k)  =  -\frac{K(k)}{\sinh(\xi K(k))} \,\, & \bigg( {\displaystyle \cosh(\xi K(k)) + \frac{1}{4} \frac{\xi K(k)}{\sinh(\xi K(k))}(k')^2 - \frac{1}{4} \cosh(\xi K(k))(k')^2 } \nonumber \\ 
& \quad + \cdots \bigg) \,\, ,
\label{eq:f1_kseries_appendix}
\end{align} 
and
\begin{align}
f_2(\xi,k)  =  -\frac{K(k)}{\sinh(\xi K(k))} \,\, & \bigg( {\displaystyle 1 + \frac{1}{4} \frac{\xi K(k) \cosh(\xi K(k))}{\sinh(\xi K(k))}(k')^2 + \frac{1}{4} \cosh^2(\xi K(k))(k')^2} \nonumber \\ 
& \quad - \frac{(k')^2}{2} + \cdots\bigg) \,\, , 
\label{eq:f2_kseries_appendix} 
\end{align} 
and
\begin{align}
f_3(\xi,k)  =  -\frac{K(k)}{\sinh(\xi K(k))} \,\, & \bigg({\displaystyle 1 + \frac{1}{4} \frac{\xi K(k) \cosh(\xi K(k))}{\sinh(\xi K(k))}(k')^2 - \frac{1}{4} \cosh^2(\xi K(k))(k')^2 } \nonumber \\ 
& \quad + \cdots  \bigg) \,\, .
\label{eq:f3_kseries_appendix}
\end{align}

\subsection{Expanding in $K(k)$}

\label{sec:K-series}

We can expand $K(k)$ as a series in $k'$,
\begin{equation}
K(k) = (\ln4 - \ln k') + \frac{1}{4} ( \ln4 - \ln k' - 1 ) (k')^2 + O(\ln k' (k')^4) \,\, ,
\end{equation}
which implies
\begin{equation}
\label{eq:expK}
e^{-2K(k)} = \frac{(k')^2}{16} + O(K(k)e^{-4K(k)}) \,\, .
\end{equation}
Using \eqref{eq:expK} in \eqref{eq:f1_kseries_appendix}, \eqref{eq:f2_kseries_appendix} and \eqref{eq:f3_kseries_appendix}, we can write $f_1$, $f_2$, and $f_3$ as series in $e^{-2K}$ to get
\begin{align}
f_1(\xi,K) = - \frac{K}{\sinh(\xi K)} \,\, & \bigg(\displaystyle{ \cosh(\xi K) + 4\left(\frac{\xi K}{\sinh(\xi K)} - \cosh(\xi K)\right) e^{-2K}} \nonumber \\ 
& \quad + O(K^2 e^{-4K}) \bigg) \,\, ,
\end{align}
and
\begin{align}
f_2(\xi,K) = - \frac{K}{\sinh(\xi K)} \,\, & \bigg({\displaystyle 1 + 4\left(\frac{\xi K\cosh(\xi K)}{\sinh(\xi K)} + \cosh^2(\xi K) - 2\right) e^{-2K} }\nonumber \\ 
& \quad + O(K^2 e^{-4K}) \bigg) \,\ ,
\end{align}
and
\begin{align}
f_3(\xi,K) = - \frac{K}{\sinh(\xi K)} \,\,  & \bigg({\displaystyle 1 + 4\left(\frac{\xi K\cosh(\xi K)}{\sinh(\xi K)} - \cosh^2(\xi K) \right) e^{-2K}} \nonumber \\ 
& \quad + O(K^2 e^{-4K}) \bigg) \,\, .
\end{align}

\section{Decoupling in the Asymptotic Limit}

\subsection{Decoupling of Zero Modes and Non-zero modes}
\label{sec:decoupling_proof}

We consider the D-strings' motion after scattering, when
\begin{equation}
\label{eq:f-approximations}
f_1 \sim 0\,\, , \quad f_2 \sim -K(t) \,\, , \quad f_3\sim 0 \,\, ,
\end{equation}
with $\dot{K}(t)$ constant. The approximations \eqref{eq:f-approximations} hold true for all $\xi$, except when $\xi$ is very close to 0 or 2, where $f_1$, $f_2$ and $f_3$ all contain poles. For now we will work with the approximations \eqref{eq:f-approximations}; we will consider the effects of the poles later on in this section.

The linearised equation of motion for $\epsilon_1$ from the Yang-Mills action is
\begin{equation}
\label{eq:epsilon1-eqn}
\ddot{\epsilon_1} - \epsilon_1'' + \epsilon_1 (f_2^2 + f_3^2) + 2f_1 (f_2\epsilon_2 + f_3\epsilon_3) = 0 \,\, ,
\end{equation} 
and the equations for $\epsilon_2$ and $\epsilon_3$ are given by cyclic permutations of \eqref{eq:epsilon1-eqn}. With the approximations \eqref{eq:f-approximations} these equations of motion become
\begin{eqnarray}
\label{eq:epsilon1-eqn-asymp}
\ddot{\epsilon}_1 - \epsilon_1'' + K^2 \epsilon_1 & = & 0 \,\, , \\
\label{eq:epsilon2-eqn-asymp}
\ddot{\epsilon}_2 - \epsilon_2''  & = & 0 \,\, , \\
\label{eq:epsilon3-eqn-asymp}
\ddot{\epsilon}_3 - \epsilon_3'' + K^2 \epsilon_3 & = & 0 \,\, .
\end{eqnarray}
Since the D-strings are far apart, $K$ is large, and \eqref{eq:epsilon1-eqn-asymp} and \eqref{eq:epsilon3-eqn-asymp} imply that
\begin{equation}
\label{eq:asymptotic-epsilon13}
\epsilon_1 = 0 \,\, , \quad \rm{and} \quad \epsilon_3 = 0 \,\, .
\end{equation}
So it seems that in the asymptotic limit the energy in the non-zero modes is entirely contained in $\epsilon_2$, which from \eqref{eq:epsilon2-eqn-asymp} takes the form of a harmonic oscillator, and has completely decoupled from the zero mode motion.

The above analysis is accurate away from the boundaries $\xi=0$ and $\xi=2$. We now consider what happens at the boundaries. Recall that the equations of motion for the full fields $\varphi_i$ derived from the Yang-Mills action, equation \eqref{eq:S_YM}, are
\begin{eqnarray}
\ddot{\varphi}_1 - \varphi''_1 + \varphi_1(\varphi_2^2 + \varphi_3^2) & = & 0  \,\, , \quad
\ddot{\varphi}_2 - \varphi''_2 + \varphi_2(\varphi_3^2 + \varphi_1^2)  =  0  \,\, , \nonumber \\
\label{eq:varphieqn-1}
\ddot{\varphi}_3 - \varphi''_3 + \varphi_3(\varphi_1^2 + \varphi_2^2) & = & 0  \,\, .
\end{eqnarray}  
Recall also that our initial condition for the $\varphi_i$ is $\varphi_i(\xi,0)=f_i(\xi,k_0)$ for some appropriate value of $k_0$. The functions $f_1(\xi,k)$, $f_2(\xi,k)$ ($f_3(\xi,k)$) are symmetric (antisymmetric) about $\xi=1$. From the equations of motion \eqref{eq:varphieqn-1} we can see that the $\varphi_i$ will be evolved in such a way that these symmetries are preserved. We therefore need only discuss the boundary at $\xi=0$, the same results will follow for the boundary at $\xi=2$ by symmetry.

Consider the initial conditions for the $\varphi_i$ at $\xi=0$. Since $\varphi(\xi,0)=f_i(\xi,k_0)$ for some $k_0$, the equations of motion \eqref{eq:varphieqn-1} give $\ddot{\varphi}_i(\xi,0)=0$. We also have, at $\xi=0$, $\dot{\varphi}_i(0,0)=0$, using the initial condition for the $\dot{\varphi}_i$ given in equation \eqref{eq:varphidot-initial-condition}, and that $df_i/dk|_{(\xi=0,k)} = 0$ for all $k$. This means that $\varphi_i(0,t)=f_i(0,k(t))$ at all times because this implies that $\dot{\varphi}_i(0,t)=\ddot{\varphi}_i(0,t)=0$ at all times, and so the $\varphi_i$ do not evolve at $\xi=0$. The $f_i$ have the following expansions for small $\xi$
\begin{eqnarray}
f_1 & = & -\frac{1}{\xi} + a_1(k) \xi + O(\xi^3) \,\, , \quad
f_2  =  -\frac{1}{\xi} + b_1(k) \xi + O(\xi^3) \,\, , \nonumber \\
\label{eq:f-series}
f_3 & = & -\frac{1}{\xi} + c_1(k) \xi + O(\xi^3) \,\, , 
\end{eqnarray}
where 
\begin{equation}
a_1(k) + b_1(k) + c_1(k) = 0 \,\, .
\end{equation}
By continuity there exists a region of small $\xi$, say $\xi<\delta$, for which the $\varphi_i$ are given by
\begin{eqnarray}
\varphi_1 & = & -\frac{1}{\xi} + a_1(k(t)) \xi + O(\xi^3)  \,\, , \quad 
\varphi_2  =  -\frac{1}{\xi} + b_1(k(t)) \xi + O(\xi^3)  \,\, , \nonumber \\
\label{eq:varphi-series}
\varphi_3 & = & -\frac{1}{\xi} + c_1(k(t)) \xi + O(\xi^3) \,\, .
\end{eqnarray}
Since $\varphi_i \to f_i$ as $\xi \to 0$, we have that $\epsilon_i \to 0$ as $\xi\to 0$, which gives us the following boundary conditions for the $\epsilon_i$
\begin{equation}
\label{eq:epsilon-bcs}
\epsilon_i (0,t) = \epsilon_i (2,t) = 0 \,\, .
\end{equation}
So we deduce that the boundary conditions \eqref{eq:epsilon-bcs} are consistent with \eqref{eq:asymptotic-epsilon13}, and $\epsilon_2$ being a harmonic oscillator.

\subsection{Energy Decoupling}
\label{sec:energy_decoupling}

We have shown in appendix \ref{sec:decoupling_proof} that the motion of the zero modes and the motion of the non-zero modes decouple when the D-strings are far apart. Therefore we also expect the energy in the non-zero modes to decouple from the energy in the zero modes in this limit. We will show here explicitly that this is the case.

The kinetic energy density is given by equation \eqref{eq:KE-density}. Substituting $\varphi_i = f_i + \epsilon_i$ we find that the coupling between the zero modes and the non-zero modes in the kinetic energy is generated by terms like
\begin{equation}
\label{eq:KE-coupling}
\int_{0}^{2} \! \dot{f}_i \dot{\epsilon}_i \, d\xi \,\, .
\end{equation}
But we have shown in appendix \ref{sec:decoupling_proof} that the non-zero $\epsilon_i$ behave like harmonic oscillators in the asymptotic limit, and the $\dot{f}_i$ are approximately constant. Then, in this limit,
\begin{equation}
\int_{0}^{2} \! \dot{f}_i \dot{\epsilon}_i \, d\xi = 0 \,\, ,
\end{equation}
and so the kinetic energy does indeed decouple. (The poles of the $f_i$ at $\xi=0$ and $\xi=2$ are fixed, as can be seen from equation \eqref{eq:varphi-series}, so that $\dot{f}_i=0$ at $\xi=0,2$. Therefore we do not need to worry about the contribution of the poles to \eqref{eq:KE-coupling}).

Next consider the potential energy. Substituting $\varphi_i = f_i + \epsilon_i$ into the potential energy density, and keeping only terms which are quadratic in $\epsilon_i$, we find that the potential energy is given by
\begin{equation}
\text{P.E.} = \frac{T}{2} \int_0^2 \left( (\epsilon_1'-f_2\epsilon_2 - f_3\epsilon_3)^2 + \{\text{cyclic perms.}\} \right) d\xi \,\, .
\end{equation}
Away from the poles the $f_i$ are given by the approximations \eqref{eq:f-approximations} in the asymptotic limit. We have also shown in appendix \ref{sec:decoupling_proof} that $\epsilon_1 = \epsilon_3 = 0$ in this limit. Then the potential energy is given by
\begin{eqnarray}
\label{eq:PE-asymp}
\text{P.E.} & = & \frac{T}{2} \int_0^{\delta} \! \left((\epsilon_2')^2 + (-f_3\epsilon_2)^2 + (-f_1\epsilon_2)^2 \right) d\xi + \frac{T}{2} \int_{\delta}^{2-\delta} \!\!\! (\epsilon_2')^2 d\xi 
\nonumber \\
&& + \frac{T}{2} \int_{2-\delta}^{2} \! \left( (\epsilon_2')^2 + (-f_3\epsilon_2)^2 + (-f_1\epsilon_2)^2 \right) d\xi \,\, ,
\end{eqnarray}
where the first and third terms in \eqref{eq:PE-asymp} take into account the behaviour of the $f_i$ near the boundaries. Here $\delta$ is a small number, chosen such that the approximations \eqref{eq:f-approximations} are accurate for $\delta < \xi < 2-\delta$. From the expressions for $f_1$, $f_2$ and $f_3$ it can be shown that $\delta \to 0$ as $k \to 1$. The series expansions \eqref{eq:f-series} for the $f_i$ imply that
\begin{equation}
f_i\epsilon_i = - \epsilon_i'(\xi)|_{\xi=0} + O(\xi^2) \,\, .
\end{equation}
So the contributions to the potential energy from the two boundary terms are given by
\begin{equation}
\label{eq:PE-boundary-terms}
\frac{3T}{2} \int_0^{\delta} \! (\epsilon_2')^2 \, d\xi + \frac{3T}{2}\int_{2-\delta}^{2} \!\!\! (\epsilon_2')^2 \, d\xi \,\, .
\end{equation}
Since these terms are finite, and $\delta \to 0$ as the D-strings get further apart, the contributions to the potential energy from \eqref{eq:PE-boundary-terms} are negligible. So the potential energy is given by
\begin{equation}
\text{P.E.} = \frac{T}{2}\int_0^2 \! (\epsilon_2')^2 d\xi \,\, ,
\end{equation}
which has also decoupled from the zero mode motion.

\bibliographystyle{utcaps}
\bibliography{D3-D1Brane_paper_num_2}

\end{document}